\documentclass[a4paper,11pt]{article}
\usepackage{jheppub} 
\usepackage{lineno}
\newcommand{\intd}{\mathrm{d}}
\newcommand{\ex}{\mathrm{e}}

\newcommand{\hp}[1]{\hphantom{#1}}
\usepackage{amsmath,epsfig}
\usepackage{amssymb,amsfonts}
\usepackage{latexsym}
\usepackage{slashed}
\usepackage{empheq}
\numberwithin{equation}{section}
\usepackage{dsfont}
\usepackage{cancel}
\usepackage{overpic}
\usepackage{subcaption}
\usepackage{enumitem}
\usepackage{subeqnarray}
\usepackage{xcolor}
\usepackage{graphicx}
\usepackage[notref,notcite]{}
\allowdisplaybreaks
\newcommand{\exclude}[1]{}
\def\nn{\nonumber}

\def\L{\mathcal{L}}

\def\d{\mathrm{d}}

\def\a#1{\alpha_{#1}}

\def\td{\tilde}

\def\2b2[#1,#2][#3,#4]{\left( \begin{array}{cc} #1 & #2 \\ #3 & #4 \end{array}
\right)}
\def\3b3[#1,#2,#3][#4,#5,#6][#7,#8,#9]{\left( \begin{array}{ccc} #1 & #2 #3 \\
#4 & #5 & #6\\#7&#8&#9\end{array} \right)}

\newcommand\fverb{\setbox\pippobox=\hbox\bgroup\verb}
\newcommand\fverbdo{\egroup\medskip\noindent%
                        \fbox{\unhbox\pippobox}\ }
\newcommand\fverbit{\egroup\item[\fbox{\unhbox\pippobox}]}

\newbox\pippobox

\def\d{\delta}

\def\6{\partial}
\def\a{\alpha}
\def\nn{\nonumber}

\def\pa{\partial}

\def\r{\rho}
\def\s{\sigma}

\def\sp{\;\;\;,\;\;\;}

\def\sq
\def\a{\alpha}

\def\l{\lambda}

\def\d{\delta}

\def\L{\Lambda}

\def\DD{{\cal D}}

\def\FF{{\cal F}}
\def\GG{{\cal G}}
\def\HH{{\cal H}}

\def\JJ{{\cal J}}

\def\MM{{\cal M}}
\def\NN{{\cal N}}
\def\OO{{\cal O}}

\def\QQ{{\cal Q}}

\usepackage{braket}

\title{$O(d,d)$ symmetric gravity and finite coupling holography}

\author{Umut Gürsoy, Pedro Vicente Marto and Edwan Préau}
\affiliation{Institute for Theoretical Physics,\\ Utrecht University, 3584 CE Utrecht, The Netherlands}

\emailAdd{p.vicentemarto@uu.nl}
\emailAdd{e.c.m.preau@uu.nl}

\abstract{We construct asymptotically AdS$_5$ black brane solutions in a theory of gravity with an infinite series of curvature corrections. The action is based on an $O(d,d)$ symmetric ansatz which has been argued to describe the classical NSNS sector of string theories. We find that, for this general class of theories, the singularity behind the horizon is not resolved by the curvature corrections. The approach to the singularity is however generically modified, being characterized by different Kasner exponents. We also show that, in the presence of a non-trivial dilaton, a slight generalization of these types of curvature corrections can generate dynamically a negative cosmological constant in the region of small coupling. This provides a mechanism through which asymptotic freedom could emerge in the hypothetical string dual of QCD.}

\begin{document}

\maketitle
\flushbottom
\newpage

\section{Introduction}

The original holographic correspondence associates a string dual to large $N$ gauge theories in the limit of strong coupling \cite{Maldacena:1997re,Itzhaki:1998dd,Gubser:1998bc,Witten:1998qj}. In this regime, the dual is described by supergravity, which is the effective description of closed string theory at the two-derivative level. Checks of the duality are abundant in this limit; see \cite{Gubser:1998bc,DHoker:2002nbb} for the most standard.

The correspondence is however expected to generalize to finite coupling, for which the dual becomes stringy, with the string length squared $\a'$ of the same order as the inverse background curvature. This is because ultimately the holographic correspondence is an example of open/closed string duality, which is more general than the short string limit. In this stringy regime, the supergravity action receives higher derivative corrections, and the corresponding solutions map to gauge theory states at finite coupling. Several checks of the correspondence exist in this regime as well, which are based on settings where the two sides of the duality share integrability properties \cite{Beisert:2010jr} (mostly $\NN=4$ super-Yang-Mills (SYM)).\footnote{Also at zero temperature a worldsheet model for the dual of free $\mathcal{N}=4$ SYM was proposed in \cite{Gaberdiel:2021qbb}.} Otherwise, this regime of the correspondence has also been tested at the level of perturbative corrections to the two-derivative supergravity action \cite{Bobev:2021qxx}.

In this work, we are interested in the finite temperature and finite coupling regime of the duality, including the limit where the gauge theory coupling vanishes (i.e. the free limit). The expected dual geometry should correspond to stringy black branes (for which integrability is a priori lost), whose horizon surface gravity still maps to the temperature of the boundary thermal state. These geometries will be solutions of an effective action of (semi-classical) string theory at the full non-local level, that is, including an infinite series of $\alpha'$ corrections. 

Gauge theory spectral functions at finite temperature are known to exhibit particular properties in the limit of vanishing coupling.  On the one hand, their singularities are branch points rather than poles, which suggests that horizons become less absorptive at weak coupling \cite{Hartnoll:2005ju,Aharony:2005bm}. On the other hand, the free correlators behave as a polynomial in the imaginary UV, $i\omega\to \pm\infty$ \cite{Hartnoll:2005ju} (with $\omega$ the frequency), which is unlike the exponential fall-off that has been argued to be a signature of the curvature singularity behind the horizon of the dual black hole \cite{Festuccia:2005pi}. This therefore suggests that, at least in the limit of vanishing coupling, the black brane duals of holographic theories should be singularity-free.  

Another interesting question related to finite coupling holography for conformal field theories (CFT), is whether a phase transition could occur in thermal states as the coupling is varied, as suggested in \cite{Li:1998kd,Gao:1998ww,Hartnoll:2005ju} for $\mathcal{N}=4$ SYM.

We will also be interested in non-conformal field theories where the coupling constant runs, which maps to a dynamical dilaton in the dual background. A concrete setting is the so-called improved holographic QCD (IHQCD) construction \cite{Gursoy:2007cb,Gursoy:2007er,Kiritsis:2009hu}, which presents the hypothetical dual of QCD as an asymptotically AdS$_5$ background of non-critical string theory. It was originally formulated in a two-derivative Einstein-dilaton theory, even though the bulk solution becomes stringy near the boundary, where the effective coupling goes to zero. This problem was not considered further in the subsequent analyses of IHQCD, since the focus was always on the strongly-coupled IR physics. Here, we would like to reconsider the UV part of the IHQCD vacuum, in a higher derivative model where the finite coupling regime can be explored. In particular, it will be interesting to test the conjecture of \cite{Gursoy:2007cb,Gursoy:2007er}, namely that the AdS curvature can be dynamically generated by higher curvature terms in the effective action for the non-critical string theory. 

\subsection*{The setup: $O(d,d)$-invariant actions}

To investigate finite $\a'$ effects in the string theories of interest, we will exploit the  $O(d,d;\mathbb{R})$ symmetry that reduced effective actions of string theory were found to exhibit for backgrounds that do not depend on $d$ coordinates. This symmetry can be seen as the manifestation of T-duality for this type of background, and it was first realised in the context of cosmological solutions to string-modified Einstein's equations in \cite{Meissner:1991zj}, where the scale factor inversion dualities exhibited in \cite{Veneziano:1991ek} were understood as $\mathbb{Z}_2^d$ transformations which can be embedded in $O(d,d)$. There, the NSNS sector of the type IIB supergravity action was shown to be expressible in a manifestly $O(d,d;\mathbb{R})$-invariant form, provided the dilaton, metric and $B$-field depend only on time. It was then proposed in \cite{Sen:1991zi} that this symmetry should hold to all orders in $\alpha^{\prime}$, based on arguments from string field theory.\footnote{An important aspect to notice is that the $O(d,d)$ transformations themselves receive $\alpha^{\prime}$ corrections; however, in \cite{Meissner:1996sa} it has been shown at first order in $\alpha^{\prime}$ that there exists a field redefinition such that these transformations take the standard uncorrected form. We emphasize that $O(d,d)$ is a symmetry of the $(10-d)$-dimensional action only, and in general it will map physical field configurations to inequivalent physical configurations. This means that the full $10$-dimensional action generically does not exhibit such a symmetry. Indeed, the recent work \cite{Hsia:2024kpi} has shown that, at order $\alpha^{\prime3}$, some field redefinitions in $(10-d)$ dimensions required for manifest $O(d,d)$ invariance of the reduced action cannot be uplifted to $10$ dimensions.}  

Recently, some progress has been made towards constructing a complete series of $\a'$ corrections to the genus-0 string effective action. The idea, which was  proposed by Hohm and Zwiebach (HZ) \cite{Hohm:2019ccp,Hohm:2019jgu}, is to constrain the form of higher derivative terms by imposing that they should be consistent with $O(d,d)$ symmetry. 
In holography, we are typically interested in type II string theories, for which it is well known that T-duality exchanges the type IIA and IIB descriptions \cite{Dai:1989ua,Dine:1988nrl,Bergshoeff:1995as}. As a result, duality does not help for constraining the RR sector of the effective action of a given type II theory. However, the NSNS sectors of the two type II theories are identical, so that useful T-duality constraints can be derived for the corresponding part of the action.  

Several interesting results have been found using the HZ action, namely the construction of non-perturbative, string frame de Sitter vacua \cite{Hohm:2019jgu,Hohm:2019ccp}, singularity resolution for $2D$ black holes \cite{Codina:2023nwz,Ying:2022xaj}\footnote{One feature that makes this possible is the target space duality in this 2D background which interchanges the horizon and singularity \cite{Dijkgraaf:1991ba,Witten:1991yr}.} and domain wall naked singularities \cite{Ying:2021xse}, and the possible resolution of the Schwarzschild black hole singularity \cite{Wu:2024eci}. 

\subsection*{Contents and outline of the paper}

Using the framework described above, we will present solutions of $O(d,d)$-invariant actions. Our method will rely on parametrizing the series of $\alpha^\prime$ corrections in the effective action as a general function of the relevant $O(d,d)$-covariant quantities.

Specializing to the case of five spacetime dimensions ($d=4$), we will first consider an asymptotically AdS$_5$ black brane ansatz with a vanishing dilaton. These are expected to be dual to thermal states in a CFT at finite coupling. We will show that, even though singularity resolution does not happen, the singularity structure of the black branes interior can generically be described by Kasner universes, with Kasner exponents that generically differ from the two-derivative Schwarzschild exponents. This modification of the Kasner exponents has been observed previously in other higher-curvature theories of gravity \cite{Bueno:2024fzg,Caceres:2024edr}, and can also be induced by the back-reaction of matter fields (for example a scalar corresponding to a relevant deformation of the dual CFT \cite{Frenkel:2020ysx,Caceres:2022smh,Caceres:2023zhl,Caceres:2023zft}). 

We also consider the possibility of a dynamical generation of an (Einstein frame) cosmological constant near the boundary of an asymptotically AdS$_5$ background with a non-trivial dilaton, based on the IHQCD construction mentioned above. We will show that this is possible with a slight generalization of the HZ ansatz, while preserving confining asymptotics in the IR limit of the geometry\footnote{This means that the IR scaling of the metric and dilaton results in an area law for Wilson loops of the dual theory.} with a vanishing 't Hooft coupling in the UV. The resulting dual field theory is thus both confining and asymptotically free.

This paper is organized as follows: in section \ref{O(d,d) actions}, we summarize the HZ construction of $O(d,d)$-symmetric actions, presenting their main features and limitations. We then extend it to actions including a generic dilaton potential, which are the more general actions considered in this paper. In section \ref{backgrounds}, we search for solutions of the equations of motion for an asymptotically AdS$_5$ black brane ansatz. We derive asymptotic constraints for the $\alpha^\prime$ corrections in field space, imposed by consistency with the zero temperature and zero coupling limits. We obtain a generic parametrization of the singularity structure in the interior of the black branes, for which example numerical solutions are presented. In section \ref{dilatonic} we turn to the study of vacuum space-times with non-trivial dilaton profiles relevant for IHQCD models. We carefully analyse the dilaton asymptotics allowed in the UV and present an explicit numerical solution. Several technical details are collected in the appendices.

\section{$O(d,d)$-invariant gravitational actions}\label{O(d,d) actions}

\label{Odd}

We introduce in this section the construction of Hohm and Zwiebach \cite{Hohm:2019jgu} for the series of higher derivative terms in the action of semi-classical string theory. It corresponds to the most general gravitational action compatible with the $O(d,d;\mathbb{R})$ symmetry of NSNS strings on geometries that depend on a single coordinate (with $D\equiv d+1$ the number of target space-time dimensions).  We will review the motivation underlying the construction of \cite{Hohm:2019jgu}, and introduce the explicit minimal form of the HZ action.  

The fundamental motivation for the HZ construction \cite{Hohm:2019jgu}, is based on the symmetries of the genus 0 NSNS type II effective action, which at quadratic order takes the following form in the string frame 
\begin{equation}
\label{IS1} I_0 = \int \intd^D x \sqrt{-G}\, \ex^{-2\phi}\left( R + 4 (\partial\phi)^2 - \frac{1}{12}H^2 \right) \, , 
\end{equation}
with $G$ the $D$-dimensional metric, $\phi$ the dilaton and $H = \intd B$, $B$ being the Kalb-Ramond two-form (or $B$-field). Focusing on solutions for which the fields depend on a single coordinate $r$ and can be written as
\begin{equation}
\label{IS2} G = 
\begin{pmatrix}
	n(r)^2 & 0 \\
	0 & g(r)
\end{pmatrix}
\sp B = 
\begin{pmatrix}
	0 & 0 \\
	0 & b(r)
\end{pmatrix}
\sp \phi = \phi(r) \, ,
\end{equation}
with $g$ and $b$ $d\times d$ matrices, the action \eqref{IS1} becomes effectively 1-dimensional
\begin{equation}
\label{IS3} I_0 = V_d \int \intd r\, n(r) \ex^{-\Phi}\left(  (\DD\Phi)^2 + \frac{1}{8} \mathrm{tr}(\DD S^2) \right) \, , 
\end{equation}
with $V_d$ the volume of the $d$-dimensional space-time transverse to $r$. The fields $\Phi(r)$ and $S(r)$ were defined as
\begin{equation}
\label{IS4} \Phi \equiv 2\phi - \frac{1}{2}\log{(-\det{g})} \sp S \equiv
 \begin{pmatrix}
 	bg^{-1} & g - bg^{-1}b \\
 	g^{-1} & -g^{-1}b
 \end{pmatrix}
 \, ,
\end{equation}
together with the derivative covariant under reparametrizations of $r$
\begin{equation}
\label{IS5} \DD \equiv \frac{1}{n(r)} \partial_r \, .
\end{equation}
The matrix $S$ in \eqref{IS4} is an element of $O(d,d;\mathbb{R})$. Thus, writing the action in terms of the fields in \eqref{IS4} makes it clear that it is invariant under an $O(d,d;\mathbb{R})$ symmetry, for which the matrix $S$ transforms by conjugation, and the field $\Phi$ is a scalar. This was first observed in  \cite{Meissner:1991zj}, and understood as an emergent enhancement of the $O(d,d;\mathbb{Z})$ symmetry from the T-duality of the full string theory \cite{Giveon:1991jj,Giveon:1994fu}. 

The $O(d,d)$ symmetry extends to the next order in $\a'$ \cite{Meissner:1991zj}, and has been argued to survive at all orders in string field theory \cite{Sen:1991zi}. Based on this observation, Hohm and Zwiebach built their series in $\a'$ by identifying all possible $O(d,d)$ invariants at each order, and using the equations of motion together with field redefinitions\footnote{Note that, importantly, the redefined fields obey the same equations of motion as the original fields, and have the same on-shell action. There is therefore no frame ambiguity for computing solutions and their on-shell actions.} to keep only a minimal set \cite{Hohm:2019jgu}. The resulting action takes the following form\footnote{Here and in the following, we removed the volume factor from the expression of the 1-dimensional action \eqref{IS3} since it does not play any role.}
\begin{equation}
\label{IS6} I = \int \intd r\, n(r) \ex^{-\Phi} \Big[ (\DD\Phi)^2 + \FF(\DD S) \Big] \, ,
\end{equation}  
where $\FF$ is a scalar function, which can be written as a formal series whose structure is constrained by $O(d,d)$ invariance
\begin{equation}
\label{IS7} \FF(\DD S) = -c_1 \mathrm{tr}[\DD S^2] - \sum_{k=2}^\infty \a' {}^{k-1} \sum_{P\in \mathrm{Part}(k,2)} c_{k,P} \prod_{m\in P} \mathrm{tr}[\DD S^{2m}] \, .
\end{equation}    
We used the same notations as in \cite{Nunez:2020hxx}, with Part$(k,2)$ the set of partitions of the integer $k$ into integers superior or equal to 2.\footnote{The number of such partitions is equal to $p(k) - p(k-1)$, with $p(k)$ the number of partitions of $k$ into positive integers.} The expression \eqref{IS7} is equivalent to the statement that $\FF$ is the most general scalar series of the single derivative $\DD S$, which involves only even powers superior or equal to 4 (with the exception of the zeroth order term $\mathrm{tr}[\DD S^2]$). The series contains an infinite number of coefficients $c_{k,P}$, labeled by an integer $k$ and an element $P$ of Part$(k,2)$, such that the order in $\a'$ is equal to $k-1$. Most of the $c_{k,P}$ are unknown, except for the first few coefficients up to $k=4$, which for type II theories are given by \cite{Gross:1986iv,Codina:2020kvj}
\begin{equation}
\label{IS8} c_1 = -\frac{1}{8} \sp c_{2,\{2\}} = 0 \sp c_{3,\{3\}} = 0 \sp c_{4,\{4\}} = -\frac{3}{2^{12}}\zeta(3) \sp c_{4,\{2,2\}} = \frac{1}{2^{12}}\zeta(3) \, .
\end{equation} 
The zeroth order term $c_1$ is directly identified from the two-derivative action \eqref{IS3}, and is the only coefficient which is the same for all string theories. The known coefficients in the case of heterotic and bosonic string theories can be found in \cite{Codina:2021cxh}. 

The form of the HZ action \eqref{IS6}-\eqref{IS7} is a very strong result, since it highly constrains the complete set of $\a'$ corrections to the genus-0 string effective action. In particular, it implies that only first derivatives of the metric appear (hence there are no Ostrogradsky ghosts), and all terms have the same exponential dependence on the dilaton. However, it is important to emphasize as well the limitations of this result:
\begin{itemize}
\item Although the structure of \eqref{IS7} is very constrained, there is still an infinite number of unknown coefficients $c_{k,P}$ that determine the higher curvature corrections. In particular the number of coefficients at a given order $k$ grows exponentially fast at large $k$, like $\ex^{\sqrt{k}}$. See Appendix \ref{AppendixA} for an example resulting in a closed form expression for the series of curvature corrections;
\item The HZ action only describes the dynamics of the NSNS fields. The RR part of the action breaks the $O(d,d)$ symmetry, so the same method cannot be applied to constrain the $\a'$ corrections involving the RR fields;
\item String loop corrections are not taken into account, which means that we are restricted to the large $N_c$ regime in holography, with no quantum gravitational effects;
\item The construction is valid only in the case where the fields depend on a single coordinate $r$. In particular, it cannot be used to compute $\a'$-complete backgrounds that contain an internal spherical factor\footnote{This is a consequence of the fact that a sphere factor cannot be preserved by T-duality, since the RR flux that supports it is itself exchanged between type IIB and type IIA. Hence, the internal space of the T-dual will have a different dimension.} (such as the $S^5$ in the dual of $\NN = 4$). This point makes the formalism a priori unsuitable for top-down holographic models. 
\end{itemize}

The last obstruction is avoided for bottom-up holographic models without internal spaces, as will be considered in the remainder of this paper. In this case, the gravitational bulk theory may be viewed as an effective description of some kind of non-critical string theory living on an asymptotically AdS background, assumed to also feature some kind of T-duality which is enhanced to $O(d,d)$ in the classical NSNS sector. This UV picture is not necessary however, and the HZ action can also be used independently, as a stringy-inspired consistent framework for investigating holographic geometries at finite $\a'$. Note that the applicability of the HZ action is still restricted to homogeneous backgrounds, which excludes in particular general fluctuations around the given background. In the holographic context, this means that the HZ action can be used to derive finite-$\a'$ homogeneous states, but not any type of correlator in those states.\footnote{That being said, homogeneous perturbations can still be analyzed with the same action, so it should be possible to compute the lowest order transport coefficients.}

To compute the homogeneous states of interest at finite $\a'$, we will want to solve the equations of motion that result from the HZ action. The latter are obtained by varying \eqref{IS6} with respect to the fields $n$, $\Phi$ and $S$
\begin{equation}
\nn E_n = (\DD\Phi)^2 - \FF(\DD S) + \mathrm{tr}[\DD S \FF'(\DD S)] = 0 \, ,
\end{equation} 
\begin{equation}
\label{IS9} E_\Phi + E_n = 2\DD^2\Phi + \mathrm{tr}[\DD S \FF'(\DD S)] = 0 \, , 
\end{equation} 
\begin{equation}
\nn E_S = \DD \QQ = 0 \, , 
\end{equation} 
with $\QQ$ the $O(d,d)$ charge, which is given by 
\begin{equation}
\label{IS10} \QQ = -2\, \ex^{-\Phi} S \FF'(\DD S) \, .
\end{equation}
This is the general form of the equations that was derived in \cite{Nunez:2020hxx}. The (contracted) Bianchi identity implies a relation between the three equations of motion \cite{Nunez:2020hxx} 
\begin{equation}
\label{IS10b} \DD E_n = \DD\Phi (E_\Phi + E_n) + \mathrm{tr}[\DD S E_S] \, .
\end{equation}
Consequently, solving $E_n$ (the zero-energy constraint) and $E_S$ (the conservation of $O(d,d)$ charge) is enough to ensure that $E_\Phi + E_n$ is also obeyed (as long as $\DD\Phi$ is non-zero). Likewise, a solution can be obtained by solving $E_S$ and $E_\Phi +E_n$, while ensuring that $E_n$ is obeyed at one point.\footnote{In holography, choices that are typically convenient would be the boundary of AdS or a horizon.} 

From now on, we will consider the case where $b(r)=0$, and the metric $g(r)$ is diagonal, so that all derivatives of $g$ and $g^{-1}$ commute. In this case, the derivative of $S$ can be written as 
\begin{equation}
\label{IS11} \DD S = 2 \HH \JJ \, , 
\end{equation}
where 
\begin{equation}
\label{IS12} \JJ \equiv 
\begin{pmatrix}
0 & g \,\,  \\
-g^{-1} & 0\,\,  
\end{pmatrix}
\sp \HH \equiv \frac{1}{2} \mathbb{I}_2 \otimes g^{-1}\DD g \, .
\end{equation}
$\HH$ is a diagonal matrix, whereas $\JJ$ obeys the following properties 
\begin{equation}
\label{IS13} \JJ^2 = - \mathbb{I}_{2d} \sp \DD\JJ = 2\HH S \, .
\end{equation}
From \eqref{IS11}-\eqref{IS13}, we deduce that 
\begin{equation}
\label{IS14} \mathrm{tr}[\DD S^{2m}] = (-4)^m \mathrm{tr}[\HH^{2m}] \, ,
\end{equation}
so that $\FF(\DD S)$ can be written as a function of $\HH$ 
\begin{equation}
\label{IS15} \FF(\DD S) = \FF_H(\HH) = 4c_1 \mathrm{tr}[\HH^2] - \sum_{k=2}^\infty (-4)^k\a' {}^{k-1} \sum_{P\in \mathrm{Part}(k,2)} c_{k,P} \prod_{m\in P} \mathrm{tr}[\HH^{2m}] \, .
\end{equation}
Note that $\FF_H(\HH)$ is itself a multivariate function of the entries of $\HH$. Using the results above \eqref{IS10}-\eqref{IS15}, the equations of motion can be written as
\begin{equation}	
\nn E_n = (\DD\Phi)^2 - \FF_H(\HH) - \ex^\Phi \mathrm{tr}[\HH \td{\QQ}] = 0 \, ,
\end{equation} 
\begin{equation}
\label{IS16} E_\Phi + E_n = 2\DD^2\Phi - \ex^\Phi \mathrm{tr}[\HH \td{\QQ}] = 0 \, , 
\end{equation} 
\begin{equation}
\nn E_S = \DD \td{\QQ} = 0 \, , 
\end{equation} 
where we defined
\begin{equation}
\nn \tilde{Q}\equiv(\s_3\otimes \mathbb{I}_d) \QQ \quad,\quad \s_3 = 
\left(\begin{array}{cc}
1 & 0 \\
0 & -1 
\end{array}\right)\, .
\end{equation}
The last equation corresponds to the conservation of the $O(d,d)$ charge. Substituting \eqref{IS10}, it implies that $\ex^{-\Phi}\FF_H'(\HH)$ is a constant.

In the following, we will allow for $O(d,d)$ breaking terms in the supergravity action \eqref{IS1}, in the form of a dilaton potential $V(\phi)$, such that \eqref{IS6} is modified to
\begin{equation}
\label{IS17} I = \int \intd r\, n(r) \ex^{-\Phi} \Big[ (\DD\Phi)^2 + \FF(\DD S) +V(\phi) \Big] \, .
\end{equation}
Such a potential could arise for example from RR fluxes. It implies the following modification to the equations of motion 
\begin{equation}	
\nn E_n = (\DD\Phi)^2 - \FF_H(\HH) - \ex^\Phi \mathrm{tr}[\HH \td{\QQ}] - V(\phi) = 0 \, ,
\end{equation} 
\begin{equation}
\label{IS18} E_\Phi + E_n = 2\DD^2\Phi - \ex^\Phi \mathrm{tr}[\HH \td{\QQ}] -\frac{1}{2}V'(\phi) = 0 \, , 
\end{equation} 
\begin{equation}
\nn E_S = \ex^\Phi\DD \td{\QQ} + \frac{1}{4}V'(\phi) = 0 \, . 
\end{equation} 
Note that the potential $V(\phi)$ actually preserves most of the $O(d,d)$ symmetry, as it breaks only the $U(1)$ subgroup corresponding to rescalings of the metric and $B$ field. The associated scalar charge  corresponds to $\mathrm{tr}(\td{\QQ})$, which is indeed non-conserved as implied by the trace of equation $E_S$ in \eqref{IS18}. On the other hand, all charges of the form  $\mathrm{tr}(\MM\td{\QQ})$ -- with $\MM$ a traceless matrix -- remain conserved. 

In this work, we consider two types of potentials: an (Einstein-frame) cosmological constant in section \ref{backgrounds}, and the improved holographic QCD (IHQCD) potential \cite{Gursoy:2007cb,Gursoy:2007er} in section \ref{IHQCD vacuum}. The latter is a non-trivial stringy-inspired potential which implies confinement in the dual theory.

\section{AdS black brane solutions and the singularity structure at finite $\a'$}

\label{backgrounds}

In this section, we investigate black brane solutions of the five-dimensional HZ action \eqref{IS17}, with a dilaton potential corresponding to a negative cosmological constant $\L$ in the Einstein frame
\begin{equation}
\label{bg1} I = \int \intd r\, n(r) \ex^{-\Phi} \Big[ (\DD\Phi)^2 + \FF(\DD S) -2 \L \ex^{-\frac{4}{3}\phi} \Big] \, . 
\end{equation}
Due to the higher curvature corrections, the black brane solutions depend on the 't Hooft coupling $\l$. These geometries can be seen as deformations of AdS$_5$-Schwarzschild, which is recovered in the limit of infinite coupling. Our main goal is to analyze how the singularity structure is affected by the curvature corrections. 

\subsection{Ansatz}

With the notations \eqref{IS2}, an appropriate ansatz for the black brane solutions of interest is given by 
\begin{equation}
\label{ans1} n(r) = \frac{\ex^{A(r)}}{\sqrt{f(r)}} \sp g(r) = \ex^{2A(r)} \mathrm{diag}\big(\!-\!\!f(r),1,1,1\big) \sp b = \phi = 0 \, ,
\end{equation}
such that the five-dimensional metric takes the form 
\begin{equation}
\label{ans1b} \intd s^2 = \ex^{2A(r)}\left(-f(r)\intd t^2 + f(r)^{-1}\intd r^2 + \intd\vec{x}^2\right) \, .
\end{equation}
Substituting this ansatz into \eqref{IS12}-\eqref{IS15}, the curvature functional $\FF(\DD S)$ in \eqref{bg1} is seen to depend in this case on two variables
\begin{equation}
\label{ans2} \FF(\DD S) = \FF\big((\DD A)^2,H_t^2\big) \sp H_t \equiv \DD A + \frac{\DD f}{2f} \, ,
\end{equation}
with the covariant derivative given by
\begin{equation}
\label{ans3} \DD = \ex^{-A(r)}\sqrt{f(r)} \pa_r \, .
\end{equation}
This implies the following form for the derivative of $\FF$ with respect to $\DD S$ 
\begin{equation}
\label{ans4} \frac{\pa \FF}{\pa \DD S} = -\frac{1}{4} \mathbb{I}_2 \otimes 
\begin{pmatrix}
\frac{\pa \FF}{\pa H_t}\Big|_{\DD A} & 0 \\
0 & \frac{1}{3}\frac{\pa \FF}{\pa \DD A}\Big|_{H_t} \mathbb{I}_3
\end{pmatrix}
\JJ \, ,
\end{equation}
with $\JJ$ defined in \eqref{IS12}, whereas the duality-invariant dilaton $\Phi$ is given in this ansatz by
\begin{equation}
\label{ans5} \Phi = - 4A - \frac{1}{2} \log{f} \, .
\end{equation}

It will be convenient to separate the quadratic part of $\FF(\DD S)$ from the rest of the curvature terms as
\begin{equation}
\label{ans6} \FF(\DD S) = -H_t^2-3\DD A^2 + \a'{}^{-1}\GG\big(\a'\DD A^2,\a'H_t^2\big) \, , 
\end{equation}
where we wrote the higher curvature terms as a general function of two variables $\GG(X,Y)$, with 
\begin{equation}
\label{ans6b} X \equiv \a'(\DD A)^2  \quad,\quad Y \equiv \a'H_t^2  \, .
\end{equation}
By definition, the behavior of $\GG$ near the origin should be such that
\begin{equation}
\label{ans7} \GG(0,0) = \pa_X\GG(0,0) = \pa_Y\GG(0,0) = 0 \, .
\end{equation}

Now the equations obeyed by the fields $A(r)$ and $f(r)$ are obtained by substituting the ansatz \eqref{ans1} into \eqref{IS18}, which gives 
\begin{equation}
\begin{aligned}
E_n&=(\mathcal{D}\Phi)^{2}-H_{t}^{2}-3(\mathcal{D}A)^{2}+2\Lambda- \a'{}^{-1}\mathcal{G}+ 2(\mathcal{D}A)^2\pa_X\GG+ 2H_{t}^2\pa_Y\GG \, ,\\
E_{\Phi}+E_n&=2\mathcal{D}^{2}\Phi-2H_{t}^{2}-6(\mathcal{D}A)^{2}-\frac{4}{3}\Lambda+ 2(\mathcal{D}A)^2\pa_X\GG + 2H_{t}^2\pa_Y\GG \, , \\
\label{Bi1} E_S&=\begin{cases}
\frac{\ex^{-4A}}{\sqrt{f}}\DD\left[\ex^{4A}\sqrt{f}\left(H_t(1-\pa_Y\GG) + \DD A(3-\pa_X\GG)\right)\right]+ \frac{8}{3}\L \, , \\
\frac{\ex^{-4A}}{\sqrt{f}}\DD\left[\ex^{4A}\sqrt{f}\left(H_t(1-\pa_Y\GG) - \DD A\left(1-\frac{1}{3}\pa_X\GG\right)\right)\right]  \, .
\end{cases}
\end{aligned}
\end{equation}
The rest of this section is devoted to analyzing the solutions to this system of equations, depending on the properties of the curvature function $\GG$. 

\subsection{AdS constraints}

We first focus on zero-temperature solutions to \eqref{Bi1}, which will give consistency constraints on the function $\GG$. Since the theories considered in this section are assumed to be conformal (the dilaton is set to 0 as in \eqref{ans1}), the dual geometry at zero temperature should be given by AdS$_5$, not only for $\a'\to 0$ but also at finite $\a'$. The bulk parameter $\a'/\ell^2$ -- with $\ell$ the AdS length such that $\Lambda = -6/\ell^2$ -- maps to the 't Hooft coupling of the dual theory, which may be written as a relation of the form $\a'/\ell^2 = h(\l)$, with $h(\l)$ a decreasing\footnote{This is implied by requiring $h(\l)$ to be one to one.} function such that $h(0) = \infty$ and $h(\infty) = 0$. 

Then, imposing that the AdS background 
\begin{equation}
\label{AdS1} A(r) = -\log\left(\frac{r}{\ell}\right) \sp f(r) = 1  \, ,
\end{equation}  
with covariant derivatives 
\begin{equation}
\label{AdS2} \DD A = -\frac{1}{\ell} \sp \DD f = 0 \, ,
\end{equation}
should be a solution of \eqref{Bi1} for any $\l$, gives the following \textit{AdS constraints} on the functions $\GG(X,Y)$ and $h(\l)$
\begin{equation}
\label{AdS3} \forall X\in\mathbb{R}_0^+, \quad \mathcal{G}(X,X) = \pa_X\mathcal{G}(X,X) =  \pa_Y\mathcal{G}(X,X) = 0\,, 
\end{equation}
\begin{equation}
\label{AdS4} \forall \l\in\mathbb{R}_0^+,\quad \l^2 h(\l)^4 = 1 \, .
\end{equation}
Note that the second condition \eqref{AdS4} implies the same relation between $\a'$ and the 't Hooft coupling as in the supergravity regime 
\begin{equation}
\label{AdS5} \frac{\a'}{\ell^2} = \l^{-\frac{1}{2}} \, .
\end{equation}

\subsection{Free limit}\label{l0_sec}

More constraints can be imposed on the function $\GG(X,Y)$ by requiring the existence of solutions in the limit where the boundary theory becomes free, $\l\to 0$. To study this limit, it is useful to use polar coordinates in field space 
\begin{equation}
\label{l01} \r \equiv \sqrt{X^2+Y^2} = \l^{-\frac{1}{2}}\ell^2\sqrt{(\DD A)^4+H_t^4} \, ,
\end{equation}
\begin{equation}
\label{l01b} \theta\equiv 
\begin{cases} 
\arctan\left(\frac{Y}{X}\right) - \frac{\pi}{4}, \quad \text{if } X>0 \\
\arctan\left(\frac{Y}{X}\right) + \frac{3\pi}{4}, \quad \text{if } X<0 \\
\end{cases}  .
\end{equation}
Note that the angle $\theta$ is defined such that it is zero on the half-line $X=Y>0$, where the AdS solutions lie. 
In these coordinates, the first two equations of motion in \eqref{Bi1} read (setting $\ell$ to 1 from now on)
\begin{equation}
	\begin{aligned}
		E_n^5&=(\mathcal{D}\Phi)^{2}-H_{t}^{2}-3(\mathcal{D}A)^{2}-12+ \l^{\frac{1}{2}}\left(-\mathcal{G}+ 2\rho \frac{\pa\mathcal{G}}{\pa \rho}\bigg|_{\theta}\right) \, ,\\
		\label{l02} E_{\Phi}^5+E_n^5&=2\mathcal{D}^{2}\Phi-2H_{t}^{2}-6(\mathcal{D}A)^{2}+8 +2\l^{\frac{1}{2}} \rho \frac{\pa\mathcal{G}}{\pa \rho}\bigg|_{\theta} \, .
	\end{aligned}
\end{equation}

In the limit $\l\to 0$, there should still exist a thermal bulk solution, with finite derivatives of order $\OO(\l^0)$. The second equation in \eqref{l02} therefore implies that $\pa_\rho \mathcal{G}$ should be finite in this limit (since $\rho = \OO(\l^{-1/2})$ from \eqref{l01}). In other words, $\mathcal{G}$ should behave linearly as $\rho$ goes to infinity
\begin{equation}
\label{l03} \mathcal{G}(\r,\theta) \underset{\r\to\infty}{\sim} C(\theta)\r \, .
\end{equation}
For vanishing coupling, the curvature corrections are thus fully captured by a single function of the compact variable $\theta\in[-3\pi/4,5\pi/4]$.\footnote{Note that the black brane solutions only explore the region $\theta\in [-\pi/4,\pi/4] \cup [3\pi/4,5\pi/4]$, where the first interval corresponds to the region outside the horizon, and the second one to the region beyond the horizon.} The AdS constraints \eqref{AdS3} further impose that $C(\theta)$ should obey
\begin{equation}
	\label{l04} C(\theta)\underset{\theta\to 0}{\sim} \theta^2 \, . 
\end{equation}

Now substituting the asymptotic behavior of $\GG$ \eqref{l03} into \eqref{l02}, the equations of motion in the $\l\to 0$ limit are found to reduce to a very simple form
\begin{equation}
	\label{l06}E_n^5=(\mathcal{D}\Phi)^{2}-H_{t}^{2}-3(\mathcal{D}A)^{2}-12+ \sqrt{(\DD A)^4+H_t^4}\, C\!\left[\arctan\left(\frac{H_t^2}{(\DD A)^2}\right)-\frac{\pi}{4}\right] \, ,
\end{equation}
\begin{equation}
	\label{l06b} (E_{\Phi}^5-E_n^5)/2 =\mathcal{D}^{2}\Phi - (\DD \Phi)^2 +16  \, .
\end{equation}
Requiring that \eqref{l06}-\eqref{l06b} admits black brane solutions with a regular horizon imposes a last condition $C\left(\theta=\frac{\pi}{4}\right)=C\left(\frac{5\pi}{4}\right)=0$, which amounts to requiring that $\GG$ vanishes at horizons.\footnote{Requiring the geometry to be smooth at the horizon actually imposes similar conditions on the $n^{\text{th}}$ derivative of $C$ at the horizon, $\partial_\theta^nC(\pi/4)=(-1)^n\partial_\theta^nC(5\pi/4)$. Since we are mostly interested in the asymptotic properties of the solutions, these conditions will not be crucial.}

\subsection{Analysis of the black branes interior}\label{rho_infty_sec}

We now address the main target of our investigation, that is the behavior of the black brane geometries behind the horizon. This analysis depends on whether the arguments \eqref{ans6b} of the curvature corrections $\GG(X,Y)$ remain finite or go to infinity at the origin. We will first consider the case where $\rho = \sqrt{X^2+Y^2}$ goes to infinity, as would happen near the Schwarzschild singularity. 

In the limit $\rho\to \infty$, the equations of motion take the same form as in the $\lambda\to 0$ limit \eqref{l06}-\eqref{l06b}, with the additional simplification that the cosmological constant becomes negligible  
\begin{equation}
\label{EI1}(\mathcal{D}\Phi)^{2}-H_{t}^{2}-3(\mathcal{D}A)^{2}+ \sqrt{(\DD A)^4+H_t^4}\, C\!\left[\arctan\left(\frac{H_t^2}{(\DD A)^2}\right)+\frac{3\pi}{4}\right] = 0 \, ,
\end{equation}
\begin{equation}
\label{EI2} \mathcal{D}^{2}\Phi - (\DD\Phi)^2 = 0 \, ,
\end{equation}
with the function $C$ defined in \eqref{l03}. To analyse the behavior near the origin, it is more convenient to use $A$ as a coordinate, for which the metric \eqref{ans1b} is written as
\begin{equation}
\label{EI2b} \intd s^2 = \frac{q(A)^2}{f(A)}\intd A^2+ \ex^{2A}\left(-f(A)\intd t^2 + \intd\vec{x}^2\right) \,,
\end{equation}
with
\begin{equation}
\label{EI2c} q(A) \equiv \frac{\ex^A}{\pa_rA} \, .
\end{equation}
The near-origin regime then corresponds to the limit $A\to-\infty$. 

Assuming a Kasner-like exponential ansatz\footnote{Imposing regular boundary conditions at the horizon, there is a unique behavior for the geometry in the interior for given curvature corrections $\GG$. Numerical solutions indicate that this behavior is exponential when $\rho$ goes to infinity in the interior.} for the fields $f(A)$ and $q(A)$   
\begin{equation}
\label{EI3} f(A) \underset{A\to-\infty}{\sim} -f_0\ex^{-\a A} \sp q(A) \underset{A\to-\infty}{\sim} -q_0 \ex^{-\a_q A} \, ,
\end{equation}
$H_t$ becomes proportional to $\DD A$
\begin{equation}
\label{EI4} H_t = \left(1-\frac{\a}{2}\right)\DD A \, ,
\end{equation}
and equation \eqref{EI2} can be written as a first order ODE for $X(A)=\l^{-1/2}(\mathcal{D}A)^2$
\begin{equation}
\label{EI5} \pa_A X + \left(8-\a\right)X = 0 \, .
\end{equation}
This equation is solved by
\begin{equation}
\label{EI6} X(A) = -X_0 \ex^{(\a-8)A} \, ,
\end{equation}
with $X_0$ a constant. This implies, from the definition of $X$, that the exponents $\a$ and $\a_q$ should be related by 
\begin{equation}
\label{EI7} \a_q = \a-4 \, .
\end{equation}
The first equation \eqref{EI1} then reduces to an algebraic condition on the Kasner exponent $\a$, which depends only on the function $C(\theta)$
\begin{equation}
\label{EI8} 6(x+1) - \sqrt{1+x^4}\, C\!\left[\arctan(x^2)+\frac{3\pi}{4}\right]=0 \sp x \equiv 1-\frac{\a}{2} \, .
\end{equation}

The interior asymptotics derived above generically differ from the Schwarzschild singularity, which is characterized by different Kasner exponents. The latter may be expressed in terms of the parameter $\a$ by writing the metric \eqref{EI2b} for $A\to -\infty$ in the Kasner form 
\begin{equation}
\label{EI9} \intd s^2 = -\intd\tau^2 + c_t^2 \tau^{2p_t}\intd t^2 + c_x^2 \tau^{2p_x}\intd \vec{x}^2 \, , 
\end{equation}
with $c_t,c_x$ two constants and $p_t,p_x$ the Kasner exponents given by
\begin{equation}
\label{EI10} p_t = \frac{2-\a}{8-\a} \quad,\quad p_x = \frac{2}{8-\a} \, .
\end{equation}
Since \eqref{EI8} does not depend on $\l$, the exponents \eqref{EI10} are the same for all finite $\lambda$. The only exception is at $\l=\infty$ for which the geometry is Schwarzschild with the exponents $p_t=-1/2$ and $p_x=1/2$, which corresponds to $\a=4$. Note that for all $\a$ the exponents \eqref{EI10} obey the first Kasner condition
\begin{equation}
\label{EI11} p_t + 3p_x = 1 \, ,
\end{equation}
but the second condition is only obeyed if $\a=4$ (i.e. for Schwarzschild exponents) 
\begin{equation}
\label{EI12} p_t^2 + 3p_x^2 = \frac{(\a-2)^2+12}{(\a-8)^2} \, .
\end{equation}
Since the Kasner conditions are implied by the Einstein equations, it is an expected feature that at least one of them should be violated in presence of higher curvature terms. The fact that the first condition is still obeyed is specific to the kind of model analyzed here, for which one equation of motion becomes $\GG$-independent in the $\rho\to\infty$ limit (see \eqref{EI2}). 

In fact, the first Kasner condition \eqref{EI11} is a consequence of \eqref{EI7}, which is itself directly implied by the $O(d,d)$ symmetry. Indeed, \eqref{EI7} can be derived from the conservation of the $O(d,d)$ charge, expressed below in \eqref{EI18}, once we impose the free limit condition \eqref{l03} on the function $\GG$.

\begin{figure}[h]
\centering
\includegraphics[scale=1]{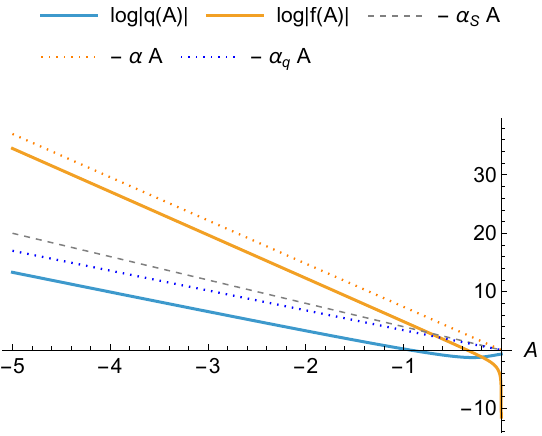} 
\caption{Numerical solution for the metric functions $f(A)$ and $q(A)$ behind the horizon ($A=0$ in the figure), for a curvature function $\GG$ chosen as in \eqref{ansG}-\eqref{C(theta)} with $c_\alpha=5$, and 't Hooft coupling $\lambda=1$. In this case, equation \eqref{EI8} has a solution at $\alpha\simeq7.40$, which agrees with the numerical result, as indicated by the dotted lines. For comparison, the dashed line indicates the asymptotic behavior of $f(A)$ near the Schwarzschild singularity, with $\a_S=4$. The numerical solution was checked to be well behaved outside the horizon as well, where it is asymptotically AdS.}
\label{fig1}
\end{figure}

\begin{figure}[h]
\centering
\includegraphics[scale=1.2]{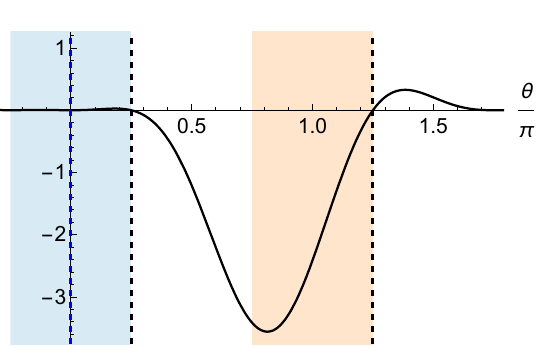} 
\caption{Plot of $C(\theta)$ in \eqref{C(theta)} for $c_\alpha=1$. The shaded regions indicate the ranges of $\theta$ that are explored by black brane solutions. The blue shaded region ($-\frac{\pi}{4}\leq\theta\leq\frac{\pi}{4}$) is the range of $\theta$ outside the horizon, whereas the orange region ($\frac{3\pi}{4}\leq\theta\leq\frac{5\pi}{4}$) corresponds to the part of the geometry that lies behind the horizon. The blue dashed line at $\theta=0$ is the AdS line, where AdS solutions lie. The black vertical lines (at $\theta=\frac{\pi}{4}$ and $\frac{5\pi}{4}$) represent the location of the horizon. It sits at a different value of $\theta$ depending on whether it is approached from the exterior or the interior.}
\label{fig2}
\end{figure}

For concreteness, we show in Figure \ref{fig1} the region behind the horizon of an explicit numerical solution, for which the Kasner exponents differ from Schwarzschild. The function $\GG$ was chosen as
\begin{equation}
\label{ansG} \GG(\rho,\theta) = C(\theta)\frac{\rho^2}{1+\rho} \, ,
\end{equation}
with\footnote{Note that, unlike the original $O(d,d)$ ansatz, this function is not analytic at $\rho=0$ (it does not admit a series expansion around $X=Y=0$). However, our conclusions are concerned with the large distance limit in field space, and analyticity at the origin does not play a role. }
\begin{equation}\label{C(theta)}
    C(\theta)=c_\alpha(1-\cos\theta)\left(\frac{\sqrt{2}}{2}-\cos\theta\right)\left(-\frac{\sqrt{2}}{2}-\sin\theta\right)\;,
\end{equation}
which is compatible with the AdS and $\l=0$ constraints derived in the previous subsections. Here $c_\alpha$ is a real constant which can be tuned to vary the amplitude of $\mathcal{G}$. This has a non-trivial effect on the solutions, since it can modify the values of $\a$ for which \eqref{EI8} is solved, and therefore the Kasner exponents in the interior. In particular, for $c_\alpha=5$ \eqref{EI8} yields $\a\simeq 7.40$. The corresponding numerical solution for the metric functions is shown in Figure \ref{fig1}, and the function $C(\theta)$ in Figure \ref{fig2}. Further details on the numerical resolution of the equations of motion and boundary conditions can be found in appendix \ref{App:num}.

\subsubsection{On the curvature singularity}\label{singularity}

Now that we have established the behavior of the solution at the origin (in the case $\rho\to\infty$), a natural question is whether the curvature singularity present at the center of the Schwarzschild solution could be resolved at finite $\a'$ by the higher curvature corrections. If this happens, in particular the Ricci scalar $R$ should remain finite, where $R$ can be expressed in terms of the metric functions $f(A)$ and $q(A)$ as
\begin{equation}
\label{EI13} 
R = 
\frac{
f'(A)\left(-9q(A) + q'(A)\right)
+ f(A)\left(-20q(A) + 8q'(A)\right)
- q(A)f''(A)
}{
q(A)^3
}
\end{equation}
However, substituting the large $\rho$ asymptotics \eqref{EI3} (taking also \eqref{EI7} into account) fixes the interior behavior of $R$ to be
\begin{equation}
\label{EI14} R \underset{A\to-\infty}{\sim} -3(\DD A)^2\left(\alpha - 4\right) \, .
\end{equation}
So we see that the singularity cannot be resolved if $\rho$ goes to infinity, since $(\DD A)^2$ goes to minus infinity in that case.\footnote{The case $\alpha = 4$, corresponding to the Schwarzschild-like behavior, is an exception for which $\DD A$ diverges but the Ricci scalar still goes to a constant, namely $R=-20/\ell^2$. This is signaled by the vanishing of the constant prefactor in \eqref{EI14} for $\alpha=4$, although one would need to compute the next order in the near-singularity behaviors \eqref{EI3} in order to recover the constant. In this case higher curvature invariants diverge at the origin (e.g. the Kretschmann scalar).} 

To give a complete proof that the singularity cannot be resolved in this kind of theories, we would need to prove that $\rho$ cannot go to a finite value at the origin ($A\to-\infty$), for every curvature function $\GG$. Although numerical results do indicate that it does not happen,\footnote{For example, for a given ansatz \eqref{ansG}, choosing a function $C(\theta)$ such that the exponent $\a$ that solves \eqref{EI8} is larger than 8 (such that the above analysis breaks down, since $\r$ would not go to infinity; see \eqref{EI6} and \eqref{EI4}), the solutions do not become regular but instead develop a curvature singularity at a finite value of $A$.} we have not been able to prove it in all generality. As we show below, it is however possible to prove the non-resolution of the singularity in two cases: 
\begin{itemize}
\item For a vanishing 't Hooft coupling $\l=0$. This means that the models analyzed in this section cannot reproduce the expectation from field theory that the singularity should be resolved in the free limit \cite{Hartnoll:2005ju}.
\item For the case where the metric functions $f(A)$ and $q(A)$ asymptote to constant values in the interior $A\to-\infty$.
\end{itemize}

We now explain why $\rho$ cannot go to a constant at $A\to-\infty$ in the case $\lambda= 0$. The first step is to prove that the blackening function $f$ remains negative behind the horizon (i.e. there is no inner horizon), which is done in Appendix \ref{AppB}. This implies that $\tilde{X}(A) \equiv f(A)/q(A)^2$ and $\tilde{Y}(A) \equiv (1+f'(A)/2f(A))^2\tilde{X}(A)$ remain negative behind the horizon.\footnote{Note that $\tilde{X},\tilde{Y}$ are the same as $X,Y$ in \eqref{ans6b}, only with the $\alpha^\prime$ factors removed.} 

The next part of the argument comes from the equation of motion \eqref{l06b}, which is independent of $\mathcal{G}(X,Y)$,  
\begin{equation}
\label{EI15} \DD^2\Phi - (\DD\Phi)^2 + 16 = 0 \, .
\end{equation}
This equation depends only on the covariant derivative of the $O(d,d)$-invariant dilaton $\DD\Phi$, which may be written in terms of the variables $\tilde{X}$ and $\tilde{Y}$ as
\begin{equation}
\label{EI16} \DD\Phi = 3i\sqrt{|\tilde{X}|} + i\sqrt{|\tilde{Y}|} \, .
\end{equation}
Assuming that $\tilde{X}$ and $\tilde{Y}$ go to constants $\tilde{X}_0,\tilde{Y}_0$ near the origin (at $A\to-\infty$), the derivative term becomes negligible in \eqref{EI15} as $A\to-\infty$, so that the condition obeyed by $\tilde{X}_0$ and $\tilde{Y}_0$ is
\begin{equation}
\label{EI17}  \Big(3\sqrt{|\tilde{X}_0|}+\sqrt{|\tilde{Y}_0|}\Big)^2 + 16 = 0 \, .
\end{equation}
Since the left-hand side is positive definite, we conclude that $\tilde{\rho} \equiv \sqrt{\tilde{X}^2 + \tilde{Y}^2}$ cannot go to a constant at the origin for $\l=0$.

Note that, for generic finite coupling, one can at least exclude the type of singularity resolution that was observed in \cite{Aguayo:2025xfi,Bueno:2025jgc}, where the metric functions $f$ and $q$ go to constants at the origin. This is seen from the conservation of the $O(d,d)$ charge (the last equation in \eqref{Bi1}), where the charge can be expressed as 
\begin{equation}
\label{EI18} \QQ_c = \ex^{4A}\frac{f(A)}{q(A)}\left[\left(1+\frac{f'(A)}{2f(A)}\right)(1-\pa_Y\GG) - 1 + \frac{1}{3}\pa_X\GG\right] \, .
\end{equation}
$\QQ_c$ can be checked to be finite, by computing it for instance at the horizon. However, if $f$ and $q$ are assumed to go to constants $f_0,q_0$ as $A\to\infty$, and taking $\GG$ to be well-behaved at finite distance in field space (in particular with $\pa_Y\GG$ finite), then \eqref{EI18} would imply that $\QQ_c$ should vanish at $A\to -\infty$.

Our results therefore indicate that the curvature singularity survives stringy corrections in the models investigated here. As shown in the previous subsection, the approach to the singularity is however modified by the curvature corrections, which is characterized by different Kasner exponents. 

Since we started from an $O(d,d)$-invariant action, one could ask whether singularity resolution could happen in other solutions related to the black branes studied above via $O(d,d)$ transformations. We check in appendix \ref{Appendix_duality} that the only transformations that leave the dilaton $\phi$ invariant (corresponding to the subgroup of $O(d,d)$ preserved by the dilaton potential in \eqref{bg1}), and do not generate a $B$-field, are (global) diffeomorphisms. This is consistent with the fact that we found a unique black brane solution for each $\l$ at $B=0$.  
Duality transformations which induce a non-trivial $B$-field might dampen the curvature growth near the singularity, or even resolve it.

\subsubsection{Kasner eons at large coupling}\label{eons}

We conclude this section by making contact with the observation made in the literature, that the near-singularity behavior of higher curvature theories is generically described by what has been called \textit{Kasner eons} \cite{Bueno:2024fzg}. This behavior is characterized by a sequence of intervals in the $A$ coordinate where the metric is well approximated by the Kasner geometry \eqref{EI9}, with different Kasner exponents for each eon. There is a change of eon each time a new term in the series of higher curvature corrections starts dominating.

\begin{figure}[h]
\begin{subfigure}{0.5\textwidth}
\includegraphics[width=1.\linewidth, height=6cm]{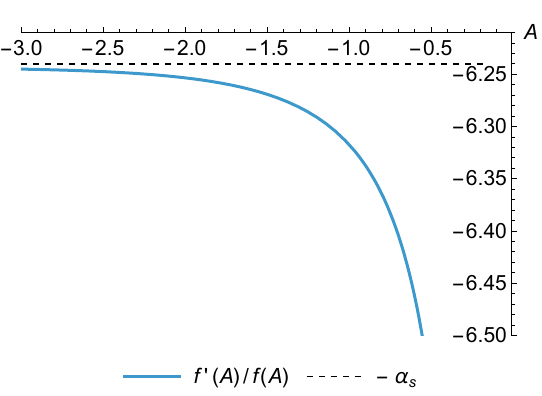} 
\caption{$\lambda=10$}
\label{fig:subim1}
\end{subfigure}
\begin{subfigure}{0.5\textwidth}
\includegraphics[width=1.\linewidth, height=6cm]{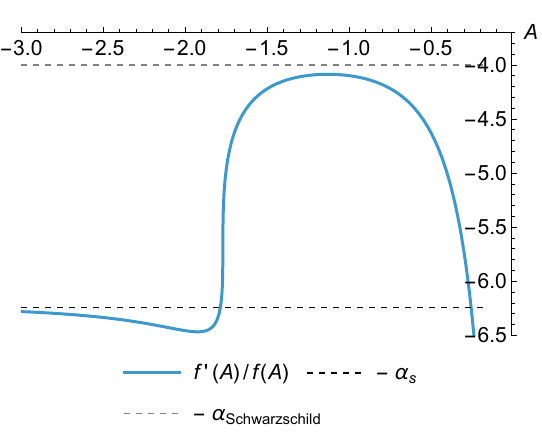}
\caption{$\lambda=10^9$}
\label{10^9}
\end{subfigure}
\caption{Effective blackening function exponent $\alpha(A)\equiv f'(A)/f(A)$ (The effective Kasner exponents depend on $\a$ as in \eqref{EI10}) for two values of the 't Hooft coupling: $\lambda=10$ (left) and $\l=10^9$ (right). The solutions are characterized by an asymptotic exponent $\alpha_s=-6.24$, obtained by setting $c_\alpha=3$ in \eqref{C(theta)}, to which they converge as the singularity is approached. This value of $\a$ is indicated as black dashed lines in the plots. The right plot includes a plateau close to the Schwarzschild exponent $\a_{\text{Schwarzschild}}=-4$, which is shown as a gray dashed line for comparison.}
\label{Kasner}
\end{figure}

In our setup, since there is only one scale $\alpha^{\prime}$ controlling the higher curvature corrections, there will be at most two Kasner eons as the singularity approached: an Einstein eon with effective Kasner exponents equal to that of the Schwarzschild singularity, and the asymptotic stringy eon with exponents \eqref{EI10}. The Einstein eon should only be visible at small enough $\a'$, that is at large enough coupling $\lambda$. 

Figure \ref{Kasner} illustrates how these expectations are realised in our setup. It shows the effective blackening factor exponent $\a(A)\equiv f'(A)/f(A)$ (defined analogously to \eqref{EI3}) for two black brane solutions at different values of the 't Hooft coupling: $\l=10$ and $\l=10^9$. Whereas a clear plateau close to the Schwarzschild exponent $\a=-4$ is visible in the $\l=10^9$ solution, there is no sign of the Einstein eon at $\l=10$. This indicates that the Kasner eon behavior is indeed visible in our setup, but only at very large values of the coupling. For most values of $\l$, the only visible Kasner behavior is that of the stringy singularity. This behaviour has been observed, for instance, in \cite{Caceres:2024edr}, where the multiple plateaus reported there relate to the multiple independent couplings present in their actions.

\begin{figure}[h]
\centering
\includegraphics[scale=1.]{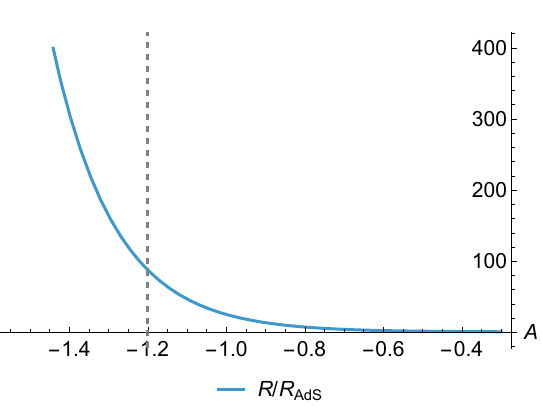}
\caption{Ricci scalar  (in units of $R_{AdS}=-20/\ell^2$) of the solution with $c_\alpha=3$ and $\lambda=10^9$. The condition $R/R_{\text{AdS}}\gtrsim100$ is verified at the point where the effective exponent $f^\prime(A)/f(A)$ is closest to the value of the Schwarzschild Kasner exponent (indicated by the gray dashed line), and beyond.}
\label{R/RAdS}
\end{figure}

That the intermediate Einstein eon emerges only at huge coupling can be understood by estimating the size of the first quadratic correction to the Einstein-Hilbert action. For a general quadratic term, the correction is expected to be subdominant when roughly $\l^{1/2} \gg (R/R_{\text{AdS}})$, with $R$ denoting the Ricci scalar of the geometry, and $R_{\text{AdS}}$ the AdS curvature. Note that $R/R_{\text{AdS}}$ is also large where the Kasner behavior arises, since it is a near-singularity behavior. If we require for instance $\l^{1/2} > 100(R/R_{\text{AdS}})$ and $(R/R_{\text{AdS}})>100$ we get roughly $\l>10^8$. For the numerical solution shown above with $c_\alpha=3$ and $\lambda=10^9$, Figure \ref{R/RAdS} shows that $R/R_{\text{AdS}}$ is indeed roughly of order $100$ where the Einstein plateau is observed, and the condition cited before is valid.

\section{Higher curvature corrections to dilatonic gravity and dynamically generated AdS}\label{dilatonic}

\label{IHQCD vacuum}

So far we have focused on an ansatz for the bulk fields of the form \eqref{ans1}, which allows for a horizon but sets the dilaton to zero. We will now focus on a different class of backgrounds with no horizons but supporting non-trivial dynamics for the dilaton. Specifically, we will focus on dilaton potentials in the improved holographic QCD (IHQCD) class, and investigate the effect of higher curvature terms on the corresponding vacuum solutions. In section \ref{action_nonc} we start by describing how the ansatz for the bulk effective action is constructed, with an $O(d,d)$ invariant form in the purely gravitational sector. After writing down the equations of motion, we examine the conditions for dynamical generation of an AdS scale by the higher curvature corrections, which corresponds to emergence of asymptotic freedom in the dual theory. We then present explicit numerical solutions for the metric and dilaton, with different kinds of asymptotics whose physical meaning is discussed.

\subsection{$O(d,d)$-invariant dilatonic gravity and the IHQCD model}
\label{action_nonc}

In this section, we are interested in theories whose bulk action is of the form 
\begin{equation}
\label{i0} I = \int \intd r\, n(r) \ex^{-\Phi} \Big[ (\DD\Phi)^2 + \FF(\DD S) +V_s(\phi) \Big] \, ,
\end{equation}
with a general dilaton potential $V_s(\phi)$. Several types of potentials have been investigated in the two-derivative Einstein-dilaton theory, and shown to generate physically consistent models.\footnote{See for example the various types of potentials investigated in the context of holographic RG studies \cite{Freedman:1999gp,Skenderis:1999mm,Kiritsis:2014kua}.} Any of these could a priori constitute a basis on which to investigate curvature corrections. Here, we will focus on a particular class of potentials that arise in the IHQCD model \cite{Gursoy:2007cb,Gursoy:2007er}. This class is of particular interest for two main reasons. First, as detailed below, the form of the IHQCD potentials is inferred from the structure expected in string theory, which is also what led us to consider $O(d,d)$-invariant higher curvature terms. And second, these potentials are such that the effective 't Hooft coupling vanishes near the AdS boundary, so that stringy effects are expected to become important there.  

We will now briefly review the IHQCD construction, before explaining how the higher curvature terms are included. Based on the arguments of \cite{Polyakov:1998ju,Kiritsis:2009hu}, the IHQCD model is an effective low-energy action for a non-critical five-dimensional string theory, which is assumed to be dual to large $N$ QCD. The field content is given by the metric $g_{MN}$, the dilaton $\phi$ and the RR 4-form $C_4$. At the two-derivative level, the IHQCD action is an Einstein-dilaton theory (plus the kinetic term for $C_4$), which takes the following form in the string frame 
\begin{equation}
\label{i1} I_s = \int\intd^5x \sqrt{-g_s}\left[\ex^{-2\phi}\left(R_s+4(\pa\phi)^2+\d c\,\a'{}^{-1}\right)-\frac{1}{2\cdot 5!}F_5^2\right] \, ,
\end{equation}
with $F_5 = \intd C_4$. The $s$ index emphasizes that the relevant metric is the string frame metric, which should also be used in tensor contractions. $\d c$ is the string frame cosmological constant, which does not vanish in a non-critical setting
\begin{equation}
\label{i2} \d c = 10 - D = 5 \, .
\end{equation} 
The equation of motion for $C_4$ is solved by 
\begin{equation}
\label{i2b} F_5 = \frac{N_c}{\ell_s} \mathrm{vol}_{5,s}\, ,
\end{equation}
with $\mathrm{vol}_{5,s}$ the string frame volume form, and $N_c$ an integration constant. It is in general proportional to the number of color, but the precise relation depends on the color brane tension \cite{Gursoy:2007cb,Kiritsis:2009hu}. Using the solution \eqref{i2b}, the 5-form may be integrated out\footnote{Note that this is not just substitution of \eqref{i2b} into \eqref{i1} (which would give the opposite sign), but rather involves writing an action which generates the same stress tensor on-shell \cite{Gursoy:2007cb}.} in \eqref{i1}, which generates instead a dilaton potential
\begin{equation}
\label{i2c} I_s =  \int\intd^5 x \sqrt{-g_s}\ex^{-2\phi}\left[R_s+4(\pa\phi)^2+\d c\a'{}^{-1}-\frac{1}{2}N_c^2\ex^{2\phi}\a'{}^{-1}\right] \, .
\end{equation}

We will now set $\a'=1$ -- which amounts to measuring all dimensionful quantities in string units -- and define the running 't Hooft coupling as
\begin{equation}
\label{i2d} \l \equiv N_c\ex^{\phi} \, .
\end{equation}
The string frame and Einstein frame metrics are then related via 
\begin{equation}
\label{i6} (g_s)_{MN} = \l^{\frac{4}{3}} g_{MN} \, , 
\end{equation}
which implies the following form for the action \eqref{i2c} in the Einstein frame 
\begin{equation}
\label{i2e} I_E = N_c^2 \int\intd r\, \sqrt{-g}\left[R- \frac{4}{3}(\pa\phi)^2 + V(\l)\right] \, ,
\end{equation}
with the Einstein frame dilaton potential given by
\begin{equation}
\label{i2f} V(\l) = \l^{\frac{4}{3}}\left(\d c - \frac{1}{2}\l^2\right) \, .
\end{equation}
In \cite{Gursoy:2007cb} (appendix B), it was argued that including higher derivative corrections to \eqref{i1} in the pure RR sector (i.e. higher powers of $F_5$) results in an action of the same form \eqref{i2e}, with a potential of the general form
\begin{equation}
\label{i2g} V_{\text{RR}}(\l) = \l^{4/3}f_{\text{RR}}(\l^2) \, ,
\end{equation}
with $f_{\text{RR}}$ admitting an analytic series expansion at small $\l$: $f_{\text{RR}} = \d c - \frac{1}{2}\l^2 + a_4\l^4 + \OO(\l^6)$. 

Potentials of the type \eqref{i2g} were found to generate consistent confining vacua for functions $f_{\text{RR}}$ obeying specific asymptotics at large $\l$ 
\begin{equation}
\label{i2h} f_{\text{RR}}(\l^2) \sim \log(\l^2)^P,\quad P\geq 0 \quad \text{or}\quad f_{\text{RR}}(\l^2) \sim (\l^2)^q\log(\l^2)^P,\quad 0\leq q\leq \frac{2}{3}(\sqrt{2}-1) \, .
\end{equation}
At small $\l$ however, the potentials \eqref{i2g} go to 0, so that there cannot be asymptotically AdS solutions with $\l$ vanishing at the boundary. This means that pure RR potentials cannot reproduce the asymptotic freedom of QCD. In \cite{Gursoy:2007cb}, it was suggested that asymptotic freedom could instead be generated by higher curvature corrections, which become large at small coupling $\l$. The main goal of our analysis in this section will be to investigate that claim.

For this we consider adding higher curvature terms to the IHQCD action \eqref{i2e}, with general RR potential \eqref{i2g}. First note that, for homogeneous backgrounds (like the IHQCD vacuum), the NSNS part of the string frame action may be written as usual in terms of $O(4,4)$ covariant quantities \eqref{IS4}-\eqref{IS5}
\begin{equation}
\label{i3} I_s = \int\intd r\, n_s(r) \ex^{-\Phi_s}\left[(\DD_s\Phi_s)^2+\frac{1}{8}\mathrm{tr}(\DD_sS_{4,s})^2+ f_{\text{RR}}(\l^2)\right] \, .
\end{equation}
We now assume that, as for critical string theories, the $O(4,4)$ symmetry extends to higher order in curvature, such that the NSNS sector of the action at all orders in $\a'$ may be written in the HZ form \eqref{IS6}
\begin{equation}
\label{i4} I_s' = \int\intd r\, n_s(r) \ex^{-\Phi_s}\left[(\DD_s\Phi_s)^2+\frac{1}{8}\mathrm{tr}(\DD_sS_{4,s})^2+ \mathcal{G}[(\DD_sS_{4,s})^2] + f_{\text{RR}}(\l^2)\right] \, .
\end{equation}
The resulting action corresponds to Einstein-dilaton improved with two series of higher derivative corrections: the pure RR terms re-packaged in the dilaton potential $f_{\text{RR}}$, and the pure NSNS higher curvature terms contained in the function $\GG$. The RR terms dominate in the IR where the dilaton $\l$ is large, whereas the higher curvature terms dominate the small $\l$ regime. Note that the action \eqref{i4} a priori does not contain mixed R-NS terms. Even though such terms never dominate at a given order in $\a'$, they may in principle have a significant contribution in the asymptotic regimes $\l\to 0$ or $\l\to\infty$. This possibility will be considered below. 

\subsection{Ansatz and equations of motion}

The NSNS ansatz relevant for the IHQCD vacuum solution takes the form
\begin{equation}
\label{i8} \intd s^2 = \ex^{2A(r)}(\intd r^2 + \intd \vec{x}^2) \sp \phi = \phi(r) \, ,
\end{equation}
which is written in the Einstein frame. This implies the following expressions for the string frame $O(d,d)$ fields 
\begin{equation}
\label{i9} n_s(r) = \ex^{A(r)}\l(r)^{\frac{2}{3}} \sp \Phi_s(r) = - 4A(r) -\frac{2}{3}\log\l(r) -2\log{N_c} \, ,
\end{equation}
\begin{equation}
\label{i10} \DD S_{4,s} = 2\left(\DD A + \frac{2\DD\l}{3\l}\right) \mathcal{J}_s \quad,\quad \DD = \ex^{-A(r)}\pa_r \, .
\end{equation}
In particular, the argument of $\mathcal{G}$ takes the form 
\begin{equation}
\label{i11} (\DD_s S_{4,s})^2 = -4\l^{-\frac{4}{3}}\left(\DD A + \frac{2\DD\l}{3\l}\right)^2 \mathbb{I}_8 \, ,
\end{equation}
so that the curvature corrections $\mathcal{G}$ can be written as a function of a single scalar variable
\begin{equation}
\label{i11b} \mathcal{G} = \mathcal{G}[\varphi] \sp \varphi \equiv \l^{-\frac{4}{3}}\left(\DD A + \frac{2\DD\l}{3\l}\right)^2 \, .
\end{equation}

The total curvature function $\mathcal{F}$ defined as in \eqref{IS17} takes the form
\begin{equation}
\mathcal{F}(\DD_s S_{4,s})=\frac{1}{8}\text{tr}{(\DD_s S_{4,s})^2} + \mathcal{G}[\varphi] \, ,
\end{equation}
with derivative
\begin{equation}
 \frac{\pa \FF}{\pa\DD_s S_{4,s}} = \frac{1}{4}\left(1-\frac{1}{4}\GG'[\varphi]\right)\DD_s S_{4,s} \, .
\end{equation}
From \eqref{IS18}, the equations of motion for the ansatz fields \eqref{i8} are then given by 
\begin{equation}\label{EOMIHQCD}
\begin{aligned}
E_n
&=\left(4\mathcal{D}A+\frac{2}{3}\DD\phi\right)^2 - 4\left(\mathcal{D}A+\frac{2}{3}\DD\phi\right)^2 - \lambda^{\frac{4}{3}}f_{\text{RR}}(\l^2)- \lambda^{\frac{4}{3}}\mathcal{G}[\varphi]+\\
&\hp{=} + 2\left(\mathcal{D}A+\frac{2}{3}\DD\phi\right)^2\GG'[\varphi]\,, \\
E_{\Phi} 
&= -2\lambda^{\frac{2}{3}}\mathcal{D}\left[\l^{-\frac{2}{3}}\left(4\mathcal{D}A+\frac{2}{3}\DD\phi\right)\right]-\left(4\mathcal{D}A+\frac{2}{3}\DD\phi\right)^2-4\left(\mathcal{D}A+\frac{2}{3}\DD\phi\right)^2 + \lambda^{\frac{4}{3}}\mathcal{G}[\varphi] + \\
&\hp{=} + \lambda^{\frac{4}{3}}\left(f_{\text{RR}}(\l^2) - \l^2 f_{\text{RR}}'(\l^2)\right)\,,\\
E_S
&= \left(\mathcal{D}^2A+\frac{2}{3}\mathcal{D}^2\phi\right) + \left(4\mathcal{D}A+\frac{2}{3}\mathcal{D}\phi\right)\left(\mathcal{D}A+\frac{2}{3}\mathcal{D}\phi+\mathcal{G}[\varphi]\right) + \mathcal{G}'[\varphi]+ \frac{1}{2} \l^2 f_{\text{RR}}'(\l^2) \, .
\end{aligned}
\end{equation}
For $\mathcal{G}=0$, we recover the equations of motion of Einstein-dilaton gravity\footnote{See, for instance, the two equations (2.11) in \cite{Gursoy:2007er}, which are combinations of equations \eqref{EOMIHQCD} at $\GG=0$.}.

In the rest of this section we analyze the UV asymptotics for solutions to these equations of motion, depending on the function $\GG$. Following this, we will present explicit numerical solutions that obey these asymptotics. 

\subsection{Looking for asymptotic freedom}
\label{asymotUV}

We are now interested to determine whether there exist curvature functions $\GG$ such that the solutions to \eqref{EOMIHQCD} admit asymptotically free solutions. That is, solutions that are asymptotically AdS in the UV ($A\to\infty$), where the coupling $\l$ also goes to 0. We will focus on cases where $\l(r)$ decays as a power-law, such that the behavior of the ansatz fields in the UV limit takes the form   
\begin{equation}
\label{i12} A(r) \underset{r\to 0}{\sim} -\log(r/\ell) \sp \l(r) \underset{r\to 0}{\sim} \l_0 r^\a \, ,
\end{equation}
with $\ell$ the AdS length (in string units) and $\l_0$ controlling the scale $\L$ of the boundary theory. Note that \eqref{i12} generically corresponds to a faster running of the dilaton compared to QCD, where the running is instead logarithmic $\l(r)\sim -\l_0/\log(r\L)$. The logarithmic running could arise in the case $\a=0$.

Assuming \eqref{i12}, the UV limit of the higher curvature terms in the equations of motion \eqref{EOMIHQCD} would go as
\begin{equation}
\label{i13} \l^{\frac{4}{3}}\mathcal{G}[\varphi]= \l^{\frac{4}{3}}\mathcal{G}\left[\l^{-\frac{4}{3}}\ell^{-2}\left(1-\frac{2}{3}\a\right)^2+\dots\right] \, ,
\end{equation}
where the dots indicate terms that are subleading in the $A\to\infty$ limit. This has a similar effect to a cosmological constant if for large $\varphi$ the function $\GG$ behaves linearly
\begin{equation}
\label{i14} \mathcal{G}[\varphi] \underset{\varphi\to\infty}{\sim} c \varphi \, ,
\end{equation}
with $c$ a constant. 
 
However, a more precise analysis of the equations of motion in the UV limit, with asymptotics \eqref{i12} and \eqref{i14}, implies that the dilaton power $\a$ obeys two incompatible equations
\begin{equation}\label{UV_AdS_eqs}
\begin{aligned}
&E_n\underset{A\to\infty}{\longrightarrow}\left(4-\frac{2}{3}\a\right)^2 + (c - 4)\left(1-\frac{2}{3}\a\right)^2 \, ,\\
&E_\Phi\underset{A\to\infty}{\longrightarrow}-4\left(4-\frac{\a^2}{9}\right) + (c-4)\left(1-\frac{2}{3}\a\right)^2
\, .
\end{aligned}
\end{equation}
Note that there is a solution $\a = 6$ for $c=4$, but in this case the equations of motion degenerate in the UV.\footnote{Concretely, this means that the subleading UV behavior of the fields is not fixed by the equations of motion.}

The conclusion of this analysis is that $O(d,d)$-invariant curvature corrections \eqref{i11b} are not sufficient to generate asymptotic freedom.\footnote{It can be checked that logarithmic or exponential runnings of the coupling instead of the power-law \eqref{i12} are also excluded.} In the next subsection, we consider a generalization of the $O(d,d)$ ansatz for which it becomes possible.

\subsubsection{Introducing $\lambda$ dependence}\label{RNS}

We now consider a more general ansatz for the higher curvature function $\mathcal{G}$, which includes independent dependence on the dilaton $\l$
\begin{equation}
\label{ld1}
\mathcal{G}[\varphi] \longrightarrow \mathcal{G}[\lambda,\varphi] \, ,
\end{equation}
with $\varphi$ still referring to the $O(d,d)$-invariant combination \eqref{i11b}. This kind of ansatz could arise from mixed RNS terms, which have the schematic form $\l^p R^q$, where $R^q$ means any contraction of $q$ curvature tensors with RR 5-fluxes $F_5$. Although there is no symmetry in the RNS sector to argue for the ansatz \eqref{ld1}, the latter is definitely a subset of all the terms that could appear, and can be seen as a simple test model for the full (NSNS, RNS and RR) higher derivative corrections of the non-critical string.  

With the modification \eqref{ld1}, the equation of motion $E_\Phi$ is modified by additional derivatives of $\GG$ with respect to $\lambda$. It now reads
\begin{equation}\label{EPHIlambda}
\begin{aligned}
    E_{\Phi} 
&= -2\lambda^{\frac{2}{3}}\mathcal{D}\left[\l^{-\frac{2}{3}}\left(4\mathcal{D}A+\frac{2}{3}\DD\phi\right)\right]-\left(4\mathcal{D}A+\frac{2}{3}\DD\phi\right)^2-4\left(\mathcal{D}A+\frac{2}{3}\DD\phi\right)^2 + \\
&\hp{=} +\lambda^{\frac{4}{3}}\left(\mathcal{G}[\lambda,\varphi] -\frac{1}{2}\lambda\partial_\lambda\GG[\lambda,\varphi]\right)+ \lambda^{\frac{4}{3}}\left(f_{\text{RR}}(\l^2) - \l^2 f_{\text{RR}}'(\l^2)\right) \, .
\end{aligned}
\end{equation}

The main interest of the generalized ansatz \eqref{ld1} for our purpose, is that it can accommodate an asymptotic behavior exactly equivalent to a cosmological constant in an asymptotically free limit \eqref{i12}, meaning
\begin{equation}
\label{ld2} \GG[\l,\varphi] \underset{\varphi=\OO(\l^{-4/3})}{\underset{\l\to 0}{\sim}} c \l^{-\frac{4}{3}} \, .
\end{equation}
Assuming this asymptotic behavior, the UV limit of the equations of motion for asymptotically free bulk fields \eqref{i12} is given by  
\begin{equation}\label{UV_AdS_eqs_RNS}
\begin{aligned}
&E_n\underset{A\to\infty}{\longrightarrow}-4\left(1-\frac{2}{3}\a\right)^2+\left(4-\frac{2}{3}\a\right)^2-\ell^2c \, ,\\
&E_\Phi\underset{A\to\infty}{\longrightarrow}- 4\left(1-\frac{2}{3}\a\right)^2-4\left(4-\frac{\a^2}{9}\right)+ \frac{5}{3}\ell^2c
\;,
\end{aligned}
\end{equation}
which has two solutions for the running of the coupling $\a$ and the AdS length $\l$
\begin{equation}\label{2sols}
\left(\a=\frac{3}{2},\ell=\frac{3}{\sqrt{c}}\right) \quad\text{and}\quad \left(\a=0,\ell=\frac{2\sqrt{3}}{\sqrt{c}}\right) \, .
\end{equation}
These two solutions correspond to qualitatively different UV RG flows: whereas the coupling runs as a power-law in the first case, it runs slower than a power-law in the second solution. Similarly to \cite{Gursoy:2007cb}, the higher order UV expansion of \eqref{UV_AdS_eqs_RNS} will typically imply a logarithmic running for the $\a=0$ solution. 

In the next subsection we present examples of numerical solutions with UV asymptotics included in \eqref{2sols}. 

\subsection{Numerical solutions}\label{num_sols}

We now look for a numerical solution of the equations of motion \eqref{EOMIHQCD}, with IR asymptotics compatible with confinement and UV asymptotics with asymptotic freedom, such that the dual field theory has QCD-like properties \cite{Gursoy:2007er,Kiritsis:2009hu}. To compute this solution, the RR potential $f_{\text{RR}}$ and the NSNS curvature function $\GG$ that appear in the action \eqref{i4} were chosen as follows. First, following the discussion around \eqref{i2h}, the RR potential is chosen to be of the form\footnote{The $1$ inside the logarithm ensures that the RR potential admits an analytic series near $\l=0$.}
\begin{equation}\label{f_RR}
f_{\text{RR}}(\lambda^2)=\log(1+\lambda^2)^P \, ,
\end{equation}
with $P>0$ to ensure confinement. The precise value of $P$ does not have an important influence on our results. To compute the solution presented below we chose $P=1/2$, which is the value for which the asymptotic glueball spectrum is linear \cite{Gursoy:2007er}, as expected in QCD. We have checked that the curvature corrections do not affect the IR asymptotics induced by the RR potential \eqref{f_RR}, so that in particular confinement is not affected. This is detailed in appendix \ref{conf_potential}.  

For the higher curvature function $\GG$, we choose a simple ansatz compatible with the asymptotically free asymptotics \eqref{ld2}\footnote{Note that this ansatz also obeys the defining conditions $\GG(\varphi=0)=\pa_\varphi\GG(\varphi=0)=0$.}  
\begin{equation}\label{GRNS}
\mathcal{G}[\lambda,\varphi]=\frac{c_0\varphi^2}{1+c_0\lambda^{\frac{4}{3}}\varphi^2/c}\;,
\end{equation}
which depends on two parameters $c$ and $c_0$. From \eqref{2sols} $c$ is directly related to the AdS length of the UV geometry (in string units), whereas the parameter $c_0$ controls the size of the curvature corrections at small $\varphi$. In particular, we found that the UV AdS geometry only arises for large enough $c_0$.

Given the curvature function and dilaton potential presented above, we solved the equation of motion $E_\Phi$ in \eqref{EPHIlambda} together with $E_n$ in \eqref{EOMIHQCD}. The problem is solved by imposing regular IR boundary conditions\footnote{The strict IR limit corresponds to $A\to-\infty$, which cannot be reached numerically. In practice, the IR boundary conditions are imposed at some finite value $A_{\text{IR}}<0$, whose absolute value should be large enough for the IR asymptotics \eqref{IR_asympt} to be a good approximation of the solution there.} and then shooting to the UV. The relevant boundary conditions and method of numerical integration are detailed in appendix \ref{App:num}. 

\begin{figure}
\begin{subfigure}{0.5\textwidth}
\includegraphics[width=1.\linewidth, height=5.5cm]{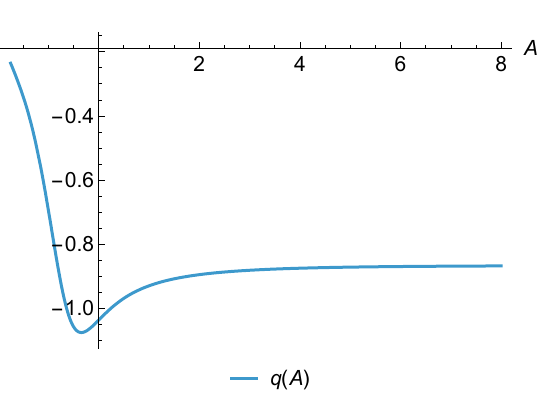} 
\end{subfigure}
\begin{subfigure}{0.5\textwidth}
\includegraphics[width=1.\linewidth, height=5.5cm]{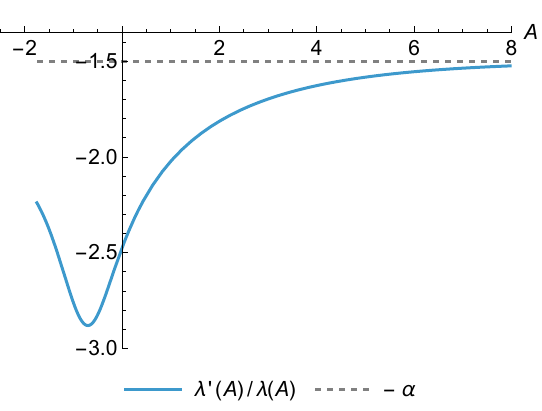}
\end{subfigure}
\caption{Metric function $q(A)$ and derivative of the dilaton $\phi'(A)=\l'(A)/\l(A)$, with $\mathcal{G}(\lambda,\varphi)$ given by \eqref{GRNS} with $c_0=50$ and $c=12$. This asymptotically free solution belongs to the first class in \eqref{2sols} - the dilaton $\phi(A)$ scales linearly with $A$ in the UV with slope $-\alpha=-\frac{3}{2}$.}
\label{aAdS}
\end{figure}

An example of a numerical background with confining IR asymptotics is shown in Figure \ref{aAdS}. We have checked that the dilaton potential is vanishing in the UV, so that the AdS scale is generated entirely by the $\alpha^\prime$ corrections contained in $\GG[\lambda,\varphi]$. The dilaton scales linearly with $A$ in the UV, with slope $-\frac{3}{2}$. This is one of the asymptotics predicted analytically in \eqref{2sols}. The asymptotic value of $q(A)$ sets the AdS length and also corresponds to the prediction from \eqref{2sols} for the value of $c$ that was chosen.

In section \ref{RNS} we noted that a second class of asymptotics is allowed for a potential like \eqref{GRNS} (the second solution in \eqref{2sols}). It corresponds to $\alpha=0$, which typically implies a logarithmic running of the dilaton in the UV. We have checked that these two classes of asymptotics are also present in the case where $\GG$ is replaced by a pure (Einstein frame) cosmological constant (i.e. the rhs of \eqref{ld2}), and that the near-boundary expansions are a priori attractive for both asymptotics. Which attractor the system chooses is, in principle, determined by dynamics. We have not found any asymptotically AdS solution corresponding to the second solution in \eqref{2sols} with $\alpha=0$. A more thorough analysis of the dynamics induced by different kinds of dilaton potentials should reveal whether these QCD-like asymptotics can also arise.

\section{Discussion}\label{outlook}

In this work, we have analyzed asymptotically AdS solutions in effective models of semi-classical string theory at finite $\a'$, which may give some insight into finite coupling holography. The backgrounds studied here are solutions to the equations of motion of an $O(d,d)$-symmetric effective action with an infinite number of $\alpha^\prime$ corrections. Parametrizing the infinite series of curvature corrections in the effective action with a bivariate function $\GG$ has allowed us to obtain on-shell solutions whose properties are controlled by $\mathcal{G}$. Our main findings were the following:

\begin{itemize}

\item In section \ref{backgrounds}, we found that the $\alpha^\prime$ corrections to asymptotically AdS$_5$ black brane solutions do not resolve the singularity behind the horizon, but do modify the near-singularity asymptotics of the metric functions. Specifically, the metric close to the singularity is still of the Kasner type, but the Kasner exponents, which can be computed from the $O(d,d)$ invariant equations of motion, differ from that of the Schwarzschild-AdS solution of Einstein gravity. An essential step in obtaining this result was recognizing that the $\l\to 0$ limit imposes a consistency condition on the asymptotic behaviour of $\GG$, namely that it should behave linearly when its arguments become large. 

\item In section \ref{IHQCD vacuum}, we studied backgrounds with no horizon and a non-trivial dilaton. We focused on asymptotically AdS spacetimes where the 't Hooft coupling vanishes in the UV, and the IR asymptotics are consistent with a dual confining theory. An $O(d,d)$-symmetric action, which in this case implies that $\GG$ is a function of a single variable, does not allow for an AdS boundary with an asymptotically free dilaton. However, asymptotic freedom can be generated when including $\lambda$ dependence in the curvature series function $\mathcal{G}$. We have thus found an explicit realization of the expectation formulated in \cite{Gursoy:2007cb} that, in an $\alpha^{\prime}$-complete theory dual to an asymptotically free gauge theory, an Einstein frame cosmological constant can be generated near the boundary by curvature corrections in the effective action. 
\end{itemize}

In most of this work we have only incorporated $O(d,d)$ symmetric corrections in the effective action coming from the NSNS sector. A consequence of our analysis is that such corrections alone are insufficient to produce regular black brane geometries. This does not mean that singularity resolution does not happen for the dual of $\mathcal{N}=4$, since our solutions are five-dimensional, and do not include RR $\a'$ corrections, which break $O(d,d)$ invariance. Our results thus suggest that the five-sphere dynamics and RR corrections play an important role to resolve the singularity in the full dual of $\NN=4$ at finite temperature and coupling.

Note that, based on the examples mentioned in the introduction \cite{Codina:2023nwz,Ying:2021xse,Ying:2022xaj,Wu:2024eci}, it could have been expected that $O(d,d)$ symmetry would be compatible with singularity resolution in our setups as well. The space of solutions studied here was however much more constrained, due to the consistency conditions that were imposed at zero temperature and zero coupling (respectively referred to as AdS constraints and free limit in the main text). These constraints make our systems interestingly more tractable, but in the end do not leave room for singularity resolution. The full solution space of $O(d,d)$ invariant actions is still largely unknown and under exploration; see for example \cite{Moitra:2025fhx} for a recent study in this direction.

Another class of higher-derivative theories of gravity that has received much attention recently are the so-called generalized quasi-topological gravities (GQTG) \cite{Bueno:2019ltp,Bueno:2019ycr}, in which several regular black hole solutions have been computed \cite{Bueno:2021krl,Bueno:2024dgm,Bueno:2025jgc,Aguayo:2025xfi}. GQTG's are higher curvature theories of gravity whose equations of motion are of second order for maximally symmetric backgrounds\footnote{The equations of motion can be higher order if covariant derivatives of the Riemann tensor are allowed in the series of higher curvature terms.}, which admit generalizations of the Schwarzschild solution that are characterized solely by the blackening function $f$. They have been argued to cover many higher curvature theories of gravity after appropriate field redefinitions \cite{Bueno:2019ltp,Bueno:2019ycr}; in particular the bosonic sector of Type IIB string theory effective action in AdS$_5$ at cubic order in $\alpha'$. It would be interesting to find the covariant field redefinitions that puts genus-0 string theory in the GQTG form for higher orders in $\alpha'$ as well. It is however difficult in the HZ approach, since $O(d,d)$ symmetry is a property of the 1-dimensional reduced action, which does not put strong constraints on the covariant general form of the action. 

In the presence of matter fields, the near-singularity Kasner behavior is known to exhibit some non-trivial dynamics, characterized by a sequence of Kasner epochs with constant  exponents, connected by abrupt transitions \cite{Hartnoll:2020fhc,Cai:2020wrp,VandeMoortel:2021gsp,Li:2023tfa,DeClerck:2023fax}. This behavior has a chaotic nature \cite{DeClerck:2023fax}, which is of a similar type that was analyzed by Belinskii, Khalatnikov and Lifshitz (BKL) \cite{Belinsky:1970ew} in the context of cosmology. Since we have observed Kasner behavior in the interior of our solutions, it is an interesting question whether BKL dynamics still arise in presence of appropriate matter fields. In particular the fields introduced in \cite{Hartnoll:2020fhc,Cai:2020wrp,VandeMoortel:2021gsp,Li:2023tfa,DeClerck:2023fax} can be coupled to the HZ action consistently, since they also depend on the holographic coordinate $r$ only.

The $O(d,d)$-invariant setups discussed here could also be used to investigate thermal states in IHQCD, by generating homogeneous black brane backgrounds with the appropriate dilaton potentials. Unlike the CFT case discussed in section \ref{backgrounds}, IHCQD models have a non-trivial phase structure at finite temperature, with a Hawking page transition between the low temperature confined phase, which is the IHQCD vacuum analyzed in this work, and the high temperature deconfined phase, corresponding to the IHCQD black branes. It would be interesting to analyze how this phase structure is modified at finite coupling (where $\ell_s/\ell_{\text{AdS}}=\mathcal{O}(1)$). In particular, it has been suggested that at weak coupling there should be an intermediate winding string gas phase \cite{Urbach:2023npi}, interpolating between the confined and the deconfined phase.

Another interesting direction building on our results would be to analyze quark-gluon plasma phenomenology within the $\a'$-complete models considered here, which include finite coupling effects. The second order nature of the equations of motion makes it a very convenient and plausible framework to investigate these effects. Stress-tensor correlators cannot be computed at finite frequency with HZ reduced actions, since the corresponding fluctuations are not invariant under time translation. However, since leading-order transport coefficients (shear and bulk viscosity) are computed from the zero-frequency limit of correlators, we expect them to be computable with an HZ ansatz for the action.\footnote{For recent work on transport coefficients in higher derivative gravity see \cite{Buchel:2023fst,Apostolidis:2025gnn,Buchel:2026xtn}.}

\section*{Acknowledgements}

We thank Nava Gaddam and David Prieto for useful discussions and comments. We also especially thank Panos Betzios and Juan Pedraza for valuable comments on the draft. This work was initiated in collaboration with Umut Gürsoy, whose insight was of major value for its development. We are thankful to him for giving this project an impetus at a critical stage. He is deeply missed. This project was supported by the Netherlands Organisation for Scientific Research (NWO) under the VICI grant VI.C.202.104. EP is supported by the European Union’s Horizon 2024 research and innovation program under the Marie Sklodowska-Curie grant agreement No 101210184.

\appendix
\section{Example of curvature corrections for a closed-form series}\label{AppendixA}

In this appendix, we give an example of coefficients $c_{n_1,n_2,P}$ such that, for the ansatz of section \ref{backgrounds}, the full series of $\alpha^\prime$ corrections in \eqref{IS15} sums to a closed form compatible with the AdS conditions \eqref{AdS3}. 

In analogy with the coefficients know up to fourth order for the tree level NSNS sector of type IIB strings in flat space \eqref{IS8}, we can build a series keeping the coefficients $c_{k,\{k\}}$ and $c_{k,\{\frac{k}{2},\frac{k}{2}\}}$ for $k$ even, and all other $c$'s set to zero. With this choice we are able to satisfy the AdS conditions. The polynomial in the action at order $\alpha^{\prime\,2p-1}$ (for $p\geq 1$) then takes the form
\begin{equation}
X_{2p}[l^{-2},l^{-2}]=2(-4)^{2p}\left(c_{2p,\{2p\}}4l^{-4p}+2\cdot c_{2p,\{p,p\}}(4l^{-2p})(4l^{-2p})\right) \, .
\end{equation}
The AdS condition $X_{k}[l^{-2},l^{-2}]=0$ implies $c_{2p,\{2p\}}=-8c_{2p,\{p,p\}}$. Similarly,
\begin{equation}
\frac{\partial X_{2p}}{\partial (\mathcal{D}A)^2}(l^{-2},l^{-2})=2(-4)^{2p}6p\left(c_{2p,\{2p\}}4l^{-2(2p-1)}+2\cdot c_{2p,\{p,p\}}(4l^{-2p+2})(4l^{-2p})\right)\text{,}
\end{equation}
which also vanishes for the relation between the coefficents written above. Note that this relation differs from that between $c_{4,\{4\}}$ and $c_{4,\{2,2\}}$ from \eqref{IS8} which, up to an overall normalization, carries a factor of $3$ instead. 

Choosing $c_{2p,{2p}}=c$ a constant, we get 
\begin{equation}
\begin{aligned}
    \mathcal{G}[(\mathcal{D}A)^2,H_t^2]&=-2c\sum_{p=1}^\infty \alpha^{\prime 2p-1}(-4)^{2p}\left[3(\mathcal{D}A)^{4p}+H_t^{4p}-\frac{1}{4}\left(3(\mathcal{D}A)^{2p}+H_t^{2p}\right)^2\right]\\
    &=-2c\sum_{p=1}^\infty \alpha^{\prime 2p-1}(-4)^{2p}\left[\frac{3}{4}(\mathcal{D}A)^{4p}+\frac{3}{4}H_t^{4p}-\frac{3}{2}(\mathcal{D}A)^{2p}H_t^{2p}\right] \, .
\end{aligned}
\end{equation}

This sums to a closed form in terms of rational functions:
\begin{equation}
\begin{aligned}    \mathcal{G}&=\frac{3}{2}c\left\{\left(\frac{1}{1+\alpha^{\prime2}(\mathcal{D}A)^4}-1\right)+\left(\frac{1}{1+\alpha^{\prime2}H_t^4}-1\right)-2\left(\frac{1}{1+\alpha^{\prime2}(\mathcal{D}A)^2H_t^2}-1\right)\right\}\\
&=6c\left\{\frac{1}{1+\alpha^{\prime2}(\mathcal{D}A)^4}+\frac{1}{1+\alpha^{\prime2}H_t^4}-2\frac{1}{1+\alpha^{\prime2}(\mathcal{D}A)^2H_t^2}\right\} \, .
\end{aligned}
\end{equation}
Note that $\mathcal{G}=0$ for $\alpha^{\prime}=0$, as it should be.

\section{Preserved subgroup of $O(d,d)$ with a dilaton potential}\label{Appendix_duality}

In the presence of an arbitrary dilaton potential $V(\phi)$ in the action \eqref{IS17}, the preserved subgroup of $O(d,d)$ duality transformations is the one which does not act on the dilaton $\phi$. To see how $\phi$ transforms under a general $O(d,d)$ transformation, one may look at the duality invariant dilaton
\begin{equation}
 \Phi=2\phi-\frac{1}{2}\log(-\det g)\,.
\end{equation}
Recall that $g$ is the $d$-dimensional part of the spacetime metric defined in  defined in \eqref{IS2}. Below we will also use $b$, the four-dimensional matrix which encodes the components of the $B$ field along the directions contained in the metric $g$. Since $\Phi$ is $O(d,d)$ invariant, the transformations that preserve $\phi$ must be the transformations that preserve $\det g$.

A generic duality transformation will affect both the metric and the $B$ field, which are packaged in the $O(d,d)$-covariant matrix $S$:
\begin{equation}\label{S}
S=\begin{pmatrix}
bg^{-1} & g-bg^{-1}b\\
g^{-1} & -g^{-1}b
\end{pmatrix}
\end{equation}
This transforms as \cite{Hohm:2019ccp}
\begin{equation}\label{Sprime}
S\,\,\to\,\,hSh^{-1}\,,\,\,\,\,\,\,\,\,\,\,\,h\in O(d,d;\mathbb{R})
\end{equation}
where a generic matrix in $O(d,d)$ may be parametrized as\footnote{By definition, elements of $O(d,d)$ are those which preserve the symplectic matrix, $h^T\eta h=\eta$.}
\begin{equation}\label{ApB4}
h=\begin{pmatrix}
A&B\\
C&D
\end{pmatrix}\,,\,\,\,\,\,\,\,\,\,\,\,\,\,h^{-1}=\begin{pmatrix}
D^T&B^T\\
C^T&A^T
\end{pmatrix}
\end{equation}
where $A^TC$ and $B^TD$ are antisymmetric and $A^TD+C^TB=\mathbb{I}_d$. 

If we restrict to backgrounds with $b=0$, the only transformations preserving $\det g$ are of the form

\begin{equation}
h=\begin{pmatrix}\label{similarity}
A&0\\
0&D\end{pmatrix}\,,\,\,\,\,\,\,\,\,\,\,\,A,D\in O(d;\mathbb{R})
\end{equation}
which act on the matrix $S$ as
\begin{equation}\label{ApB6}
S\,\,\to\,\,\begin{pmatrix}
0 & AgD^{-1}\\
Dg^{-1}A^{-1} & 0
\end{pmatrix}
\end{equation}
so that the metric transforms as $g\to AgD^{-1}=AgA^T$, which is a global change of coordinates.\footnote{If we act with this transformation on a background with non-zero $b$, then the metric transforms in the same way and $b\to AbD^{-1}$.}  The condition that $A$ and $D$ are in $O(d)$ comes from the requirement that the determinant of $g$ be preserved. 

Transformations which generate a non-trivial $B$ field will generically affect the metric and curvature\footnote{An exception is the simple example with $A=D=\mathbb{I}_d$, $C=0$ and $B=-B^T$, which induces a constant $B$ field, for which the metric is unchanged.}, but they lie outside the class of solutions that we are interested in. 

\section{Absence of inner horizon}

\label{AppB}

In this appendix, we prove that the black brane solutions of section \ref{backgrounds} do not contain inner horizon, in the case of zero coupling $\l=0$. The equations of motion for this case are given by \eqref{l06}-\eqref{l06b}, together with the conservation and non-conservation equations for the $O(d,d)$ charges in \eqref{Bi1}
\begin{equation}
\label{B1}(\mathcal{D}\Phi)^{2}-H_{t}^{2}-3(\mathcal{D}A)^{2}-12+ \sqrt{(\DD A)^4+H_t^4}\, C\!\left[\arctan\left(\frac{H_t^2}{(\DD A)^2}\right)-\frac{\pi}{4}\right] = 0 \, ,
\end{equation}
\begin{equation}
\label{B2} \mathcal{D}^{2}\Phi - (\DD \Phi)^2 +16 = 0  \, ,
\end{equation}
\begin{equation}
\label{B3} \DD Q_c = 0 \quad,\quad \DD Q_{nc} = -\frac{8}{3}\L\ex^{-\Phi}\, ,
\end{equation}
where the charges are expressed in $A$ coordinates as 
\begin{equation}
\label{B4} Q_c = \ex^{4A}\frac{f(A)}{q(A)}\left[\left(1+\frac{f'(A)}{2f(A)}\right)(1-\pa_Y\GG) - 1 + \frac{1}{3}\pa_X\GG\right] \, ,
\end{equation}
\begin{equation}
\label{B5} Q_{nc} = \ex^{4A}\frac{f(A)}{q(A)}\left[\left(1+\frac{f'(A)}{2f(A)}\right)(1-\pa_Y\GG) + 3 - \pa_X\GG\right] \, .
\end{equation}

At $\l=0$ the curvature functional $\GG(X,Y)$ takes the asymptotic form \eqref{l03} 
\begin{equation}
\label{B6} \GG(X,Y) = \rho(X,Y) C(\theta(X,Y))\,,
\end{equation}
with $\rho$ and $\theta$ the polar coordinates \eqref{l01}-\eqref{l01b}. This implies the following expressions for the derivatives of $\GG$ in terms of $C(\theta)$
\begin{equation}
\label{B7} \pa_X \GG = \cos\left(\theta+\frac{\pi}{4}\right)C(\theta) - \sin\left(\theta+\frac{\pi}{4}\right)C'(\theta) \, ,
\end{equation}
\begin{equation}
\label{B8} \pa_Y \GG = \sin\left(\theta+\frac{\pi}{4}\right)C(\theta) + \cos\left(\theta+\frac{\pi}{4}\right)C'(\theta) \, .
\end{equation}

Now let us assume that there is an inner horizon at some $A_*>A_H$ (i.e. $f(A_*) = 0$), with $A_H$ the location of the (outer) horizon. We will take $A_*$ to be the location of the first inner horizon, such that $f(A)<0$ for $A_H<A<A_*$. The equality of the conserved charge \eqref{B4} at the two horizons then implies\footnote{Note that the angle $\theta$ is the same at the two horizons $A=A_H^+$ and $A=A_*^-$: $\theta_H = 5\pi/4$.} 
\begin{equation}
\label{B9} \ex^{4A_H}\frac{f'(A_H)}{2q(A_H)} = \ex^{4A_*}\frac{f'(A_*)}{2q(A_*)} \, .
\end{equation}
In particular, the sign of $f'(A)/q(A)$ should be the same at the two horizons, so $f'(A_*)/q(A_*)$ should be positive. Therefore, if $q(A_*)$ remains negative, then $f'(A_*)$ should also be negative, which is in contradiction with $f(A)$ approaching 0 from below at $A=A_*$. 

To complete the proof, we now need to show that $q(A)$ actually remains negative between the two horizons. This means showing that there is no point $A_0\in(A_H,A_*)$ such that $q(A_0)=0$. Assuming that such a point exists, equation \eqref{B1} together with the condition that the $O(d,d)$ charges \eqref{B4}-\eqref{B5} should remain finite impose three conditions on $x(A)\equiv 1+f'(A)/(2f(A))$ 
\begin{equation}
\label{B10} 6\big(x(A_0)+1\big) - \sqrt{1+x(A_0)^4}C\big(\theta\big[x(A_0)\big]\big) = 0 \, ,
\end{equation}
\begin{equation}
\label{B11} x(A_0)\Big(1-\pa_Y\GG\big(\theta\big[x(A_0)\big]\big)\Big)- 1+ \frac{1}{3}\pa_X\GG\big(\theta\big[x(A_0)\big]\big) = 0 \, ,
\end{equation}
\begin{equation}
\label{B12} x(A_0)\Big(1-\pa_Y\GG\big(\theta\big[x(A_0)\big]\big)\Big)+ 3- \pa_X\GG\big(\theta\big[x(A_0)\big]\big) = 0 \, ,
\end{equation}
where, from \eqref{l01b}, the angle $\theta$ is expressed in terms of $x(A_0)$ as
\begin{equation}
\label{B13} \theta\big[x(A_0)\big] = \arctan\big[x(A_0)^2\big] + \frac{3\pi}{4} \, ,
\end{equation}
and the derivatives of $\GG$ are given in \eqref{B7}-\eqref{B8}. 

The system \eqref{B10}-\eqref{B12} has a unique solution for $x(A_0),C\big(\theta\big[x(A_0)\big]\big)$ and $C'\big(\theta\big[x(A_0)\big]\big)$
\begin{equation}
 x(A_0) = -3 \quad,\quad C\big(\theta\big[x(A_0)\big]\big) = -6\sqrt{\frac{2}{41}} \quad,\quad C'\big(\theta\big[x(A_0)\big]\big) = 13\sqrt{\frac{2}{41}} \, .
\end{equation}
In other words, \eqref{B10}-\eqref{B12} generically cannot be obeyed, and the only exception is when $C(\theta)$ and $C'(\theta)$ take the specific values \eqref{B13} at the correpsonding angle $\theta\big[x(A_0)\big] = \arctan(9)+3\pi/4$. However, even in this case, it can be checked that equation \eqref{B2} together with the derivative of \eqref{B1} constrain the derivative of $x(A)$ and $q(A)$ to vanish as 
\begin{equation}
\label{B14} x'(A_0) = \OO(q(A_0)) \quad,\quad q'(A_0) = \OO(q(A_0)) \, .
\end{equation}
The Ricci scalar \eqref{EI13} is then found to diverge as $R\propto q(A_0)^{-2}$, so that there is a curvature singularity at $A_0$. 

The conclusion of the above analysis is that for almost all functions $\GG(X,Y)$, $q(A)$ cannot go to zero between the two horizons, so that it remains negative. There is a family of functions for which it might be possible, but the corresponding point would then correspond to a curvature singularity. According to our arguments above, we can therefore conclude that black brane solutions at $\l=0$ do not admit inner horizons. Note that, since $\rho$ goes to infinity at the special points relevant for our analysis (i.e. $A_H,A_*$ and $A_0$), we expect the result to generalize to finite $\l$. 

\section{IR confining asymptotics} \label{conf_potential}

In this work we have seen different scenarios where $\alpha^\prime$ corrections can dynamically generate a cosmological constant near the AdS boundary. In this appendix, we check that the curvature corrections cannot become dominant in a confining background, thus the IR physics is not affected. This is desirable, since we don't want to completely reconsider the model, but it also means that the IR singularity at the bottom of the confining geometry remains.\footnote{As mentioned in \cite{Gursoy:2007cb}, this singularity is of the good kind according to Gubser's criteria \cite{Gubser:2000nd}.}

We will be interested in the scenario of particular physical relevance for which the dilaton potential and behaviour of the fields in the IR results in confinement, that is, an area law for the action of a fundamental string attached at the boundary and extending into the bulk. This was studied in \cite{Gursoy:2007er}, where the IR asymptotics of the metric functions and the dilaton potential required for such an area law were classified. 

Take for instance geometries which, in the conformal frame\footnote{This is defined as the frame where the metric takes the form $ds^2=e^{2A(r)}[dr^2+\eta_{ij}dx^idx^j]$. The AdS boundary is at $r\to 0$ while the bottom of the geometry is at $r\to\infty$.} as $r\to\infty$, take the form
\begin{equation}\label{IR_asympt}
e^{A(r)}\sim e^{-\left(\frac{r}{\ell}\right)^{\alpha}}\text{,}\,\,\,\,\,\,\,\,\,\, \lambda(r)\sim e^{\frac{3}{2}\left(\frac{r}{\ell}\right)^{\alpha}}\left(\frac{r}{\ell}\right)^{\frac{3}{4}(\alpha-1)}\text{.}
\end{equation}
These are confining for $\alpha\geq 1$ if the Einstein frame dilaton potential behaves asymptotically at large $\l$ as
\begin{equation}\label{Vpot}
V(\lambda)\sim\lambda^{\frac{4}{3}}(\log\lambda)^{\frac{\alpha-1}{\alpha}}=\lambda^{\frac{4}{3}}\left(\frac{3}{2}\left(\frac{r}{\ell}\right)^{\alpha}+\frac{3}{4}(\alpha-1)\frac{r}{\ell}\right)^{\frac{\alpha-1}{\alpha}} \, .
\end{equation}
Now, the variable $\varphi$ defined in \eqref{i11b} that enters the $O(d,d)$ series of higher derivative terms is given by
\begin{equation}
\begin{aligned}
\varphi\,&=\,\lambda^{-\frac{4}{3}}e^{-2A(r)}\left(\partial_rA(r)+\frac{2}{3}\frac{\partial_r\lambda(r)}{\lambda(r)}\right)^2\\
\,&=\,\frac{(\alpha-1)^2}{4}\lambda^{-\frac{4}{3}}e^{2\left(\frac{r}{\ell}\right)^{\alpha}}r^{-2}
\,.
\end{aligned}
\end{equation}
This quantity decays as $r^{-(\alpha+1)}$ when $r\to\infty$, so the argument of the series functional $\mathcal{G}$ goes to zero in the IR. The contribution of the NSNS curvature corrections to the action in the IR limit is then 
\begin{equation}\label{Gexpansion}
\begin{aligned}
\lambda^{\frac{4}{3}}\mathcal{G}[\varphi]\,&=\,\lambda^{\frac{4}{3}}\left[\frac{1}{2}\mathcal{G}''[0]\varphi^2+\mathcal{O}(\varphi^3)\right]\\
\,&=\,e^{2\left(\frac{r}{\ell}\right)^{\alpha}}\left(\frac{r}{\ell}\right)^{(\alpha-1)}\left[\mathcal{G}^{(2)}\frac{(\alpha-1)^2}{32}r^{-2(\alpha+1)}+\mathcal{O}(\varphi^3)\right]\\
\,&\sim\, e^{2\left(\frac{r}{\ell}\right)^{\alpha}}r^{-(\alpha+3)}\,.
\end{aligned}
\end{equation}
Comparing with \eqref{Vpot}, the ratio between the curvature corrections and the dilaton potential in this limit scales as
\begin{equation}
\frac{\lambda^{\frac{4}{3}}\mathcal{G}[\varphi]}{V(\lambda)}\underset{r\to\infty}{\,\sim}\,r^{-(3\alpha+4)}
\end{equation}
We thus see that curvature corrections are subleading with respect to RR terms in the (Einstein frame) action which become dominant in the limit where the dilaton grows.

This result still holds for the deformed ansatz with dilaton dependence introduced in section \ref{RNS}. From \eqref{Gexpansion} it is seen that the curvature corrections for the RNS-like ansatz \eqref{GRNS} indeed take the form 
\begin{equation}
\lambda^{\frac{4}{3}}\GG[\l,\varphi]\underset{r\to \infty}{=}c+\mathcal{O}(e^{-2\left(\frac{r}{\ell}\right)}r^{(\alpha+3)})\,,
\end{equation}
which tends to a constant, and is therefore even more subleading than \eqref{Gexpansion} with respect to \eqref{Vpot}.

The analysis above proves that the type of $\alpha^{\prime}$ corrections considered in our setup, although not completely negligible, are insufficient to generate a dilaton potential which leads to confinement, so that it must be generated by corrections to the effective action from a different sector. This role is played instead by the RR flux, which also dominates relative to R-NS contributions (mixed curvature-flux higher derivative terms). 

As mentioned above, the fact that pure curvature corrections are negligible in the IR compared with the RR sector implies that a resolution of the curvature singularity at the origin of the IHQCD vacuum metric will not happen in our setup. 

\section{Details on numerical solutions}\label{App:num}

We solve a system of two (independent) differential equations; typically we choose $E_n$ and $E_{\Phi}$. Due to the Bianchi identity \eqref{IS10b}, these two equations are enough to obtain the complete solution to the equations of motion. There are two distinct cases to discuss, that of black branes studied in section \ref{backgrounds} and that of vacuum dilatonic backgrounds studied in section \ref{dilatonic}.

\subsection{Black brane solutions}

When imposing the black brane ansatz \eqref{ans1b}, we solve for the functions $f(A)$ and $q(A)$. The relevant equations of motion are
\begin{equation}
\label{EOMbb}
\begin{aligned}
E_n^5&=(\mathcal{D}\Phi)^{2}-H_{t}^{2}-3(\mathcal{D}A)^{2}-12+ \l^{\frac{1}{2}}\left(-\mathcal{G}+ 2\rho \frac{\pa\mathcal{G}}{\pa \rho}\bigg|_{\theta}\right) \, ,\\
E_{\Phi}^5+E_n^5&=2\mathcal{D}^{2}\Phi-2H_{t}^{2}-6(\mathcal{D}A)^{2}+8 +2\l^{\frac{1}{2}} \rho \frac{\pa\mathcal{G}}{\pa \rho}\bigg|_{\theta} \, ,
\end{aligned}
\end{equation}
where we recall that
\begin{equation}
\DD = \frac{\sqrt{f(A)}}{q(A)}\partial_A \quad,\quad H_t = \frac{\sqrt{f(A)}}{q(A)}\left(1+\frac{f'(A)}{2f(A)}\right) \quad,\quad \Phi = -4A - \frac{1}{2}\log f(A) \, . 
\end{equation}
Equations \eqref{EOMbb} are coupled ODE's of second order for $f$ and of first order for $q$, that can be solved with the NDSolve function of Mathematica. The numerical solutions to \eqref{EOMbb} are obtained by shooting from the horizon at $A=0$ to an arbitrary positive/negative value of $A$, according to whether we look for the solution in the exterior/interior. We therefore need to provide two boundary conditions for $f$ at the horizon and one for $q$. Namely: $f(A=0)=0$, which defines the location of the horizon, and $\partial_Af(A)|_{A=0}=1$, which fixes the normalization of the horizon temperature. This latter value can be freely tuned. Once these conditions are imposed on $f$, the near-horizon equations of motion do not leave more freedom on the value of $q$ at the horizon, which has to obey $q(A=0)=-\frac{1}{2}\sqrt{f^\prime(A=0)}$.  

We found that the numerical integration with NDSolve behaved better when inputting the system of equations \eqref{EOMbb} directly in the form  
\begin{equation}
\label{E2} f''(A) = E_f[f'(A),f(A),q(A)] \quad,\quad q'(A) = E_q[f'(A),f(A),q(A)] \, .
\end{equation}

The equations of motion \eqref{EOMbb} depend on the choice of function $\GG$, as well as the parameter $\l$. Although we only investigated a small number of choices, our numerical results indicate that, for fixed asymptotics, different choices of $\GG$ and $\l$ do not induce major qualitative differences on the solutions.

\subsection{Dilatonic solutions}

We now turn to the case of the dilatonic backgrounds of section \ref{dilatonic}. We solve the equations $E_n$ and $E_\Phi$ for $\lambda(A)$ and $q(A)$, 
\begin{equation}\label{EOMd}
\begin{aligned}
E_n
&=\left(4\mathcal{D}A+\frac{2}{3}\DD\phi\right)^2 - 4\left(\mathcal{D}A+\frac{2}{3}\DD\phi\right)^2 - \lambda^{\frac{4}{3}}f_{\text{RR}}(\l^2)- \lambda^{\frac{4}{3}}\mathcal{G}[\varphi]+\\
&\hp{=} + 2\left(\mathcal{D}A+\frac{2}{3}\DD\phi\right)^2\GG'[\varphi]\,,\\
    E_{\Phi} 
&= -2\lambda^{\frac{2}{3}}\mathcal{D}\left[\l^{-\frac{2}{3}}\left(4\mathcal{D}A+\frac{2}{3}\DD\phi\right)\right]-\left(4\mathcal{D}A+\frac{2}{3}\DD\phi\right)^2-4\left(\mathcal{D}A+\frac{2}{3}\DD\phi\right)^2 + \\
&\hp{=} +\lambda^{\frac{4}{3}}\left(\mathcal{G}[\lambda,\varphi] -\frac{1}{2}\lambda\partial_\lambda\GG[\lambda,\varphi]\right)+ \lambda^{\frac{4}{3}}\left(f_{\text{RR}}(\l^2) - \l^2 f_{\text{RR}}'(\l^2)\right) \, ,
\end{aligned}
\end{equation}
with 
\begin{equation}
\DD\phi = \frac{\DD\l}{\l} \quad,\quad \DD = q(A)^{-1}\pa_A \, .
\end{equation}
The RR potential $f_{RR}$ is chosen as in \eqref{f_RR}. Equations \eqref{EOMd} are of second order in $\lambda$ and first order in $q$. We solve these again with NDSolve, by shooting from an IR cutoff $A_{min}<0$, where we impose that the metric function $A(r)$ and dilaton $\l(r)$ behave according to the confining asymptotics \eqref{IR_asympt} 
\begin{equation}\label{bc1}
A(r)\underset{r\to\infty}{\sim}-\left(\frac{r}{\ell}\right)^\alpha\,,
\end{equation}
\begin{equation}\label{bc2}
\phi(r)\underset{r\to\infty}{\sim}\frac{3}{2}\left(\frac{r}{\ell}\right)^\alpha+\frac{3}{4}(\a-1)\log\left(\frac{r}{\ell}\right)\,.
\end{equation}
Since numerically we express our system of equations in terms of the independent radial variable $A$ instead of $r$, we need to express these boundary conditions in terms of the IR cut-off $A_{min}$. Redefining the radial variables from $r$ to $A$ according to $\frac{e^A}{\partial_rA}=q(A)$ gives
\begin{equation}
\phi(A_{min})=-\frac{3}{2}A_{min}+\frac{3}{4} \log\left|\alpha A_{min}^\frac{\alpha-1}{\alpha}\right|\,,\,\,\,\,\,\,\,\,\,\partial_A\phi(A_{min})=-\frac{3}{2}+\frac{3}{4}\frac{\alpha-1}{\alpha |A_{min}|} \, ,
\end{equation}
and
\begin{equation}
q(A_{\min})=-\frac{e^{A_{min}}}{\alpha A_{min}^\frac{\alpha-1}{\alpha}}\,.
\end{equation}

Note that the IR cutoff $|A_{min}|$ should be large enough for \eqref{bc1} and \eqref{bc2} to be valid in the IR regime. The integration is terminated at a UV cutoff $A_{max}$, that should be large enough for the asymptotics \eqref{2sols} to be observed. As stated in the main text, we have verified numerically that our equations of motion are satisfied, up to numerical tolerance, throughout the regime of integration.

A final comment on the numerical resolution of the equations is that, once the asymptotics of the solution are fixed, there is no major qualitative difference between solutions produced with different functions $\mathcal{G}$. However, as noted in section \ref{RNS}, there is a second class of AdS asymptotics, in principle allowed by the equations of motion, that we have not observed numerically. It is plausible that these asymptotics can be obtained with sufficiently different choices of function $\GG$.

Also, in this case as well, we found it more efficient to input the equations \eqref{EOMd} into NDSolve in the form 
\begin{equation}
\label{Ef} \l''(A) = E_\l[\l'(A),\l(A),q(A)] \quad,\quad q'(A) = E_q[\l'(A),\l(A),q(A)] \, .
\end{equation}

\bibliography{biblio}

@article{Bueno:2021krl,
    author = "Bueno, Pablo and Cano, Pablo A. and Moreno, Javier and van der Velde, Guido",
    title = "{Regular black holes in three dimensions}",
    eprint = "2104.10172",
    archivePrefix = "arXiv",
    primaryClass = "gr-qc",
    doi = "10.1103/PhysRevD.104.L021501",
    journal = "Phys. Rev. D",
    volume = "104",
    number = "2",
    pages = "L021501",
    year = "2021"
}

@article{Bueno:2024dgm,
    author = "Bueno, Pablo and Cano, Pablo A. and Hennigar, Robie A.",
    title = "{Regular black holes from pure gravity}",
    eprint = "2403.04827",
    archivePrefix = "arXiv",
    primaryClass = "gr-qc",
    doi = "10.1016/j.physletb.2025.139260",
    journal = "Phys. Lett. B",
    volume = "861",
    pages = "139260",
    year = "2025"
}

@article{Bueno:2019ycr,
    author = "Bueno, Pablo and Cano, Pablo A. and Hennigar, Robie A.",
    title = "{(Generalized) quasi-topological gravities at all orders}",
    eprint = "1909.07983",
    archivePrefix = "arXiv",
    primaryClass = "hep-th",
    reportNumber = "IFT-UAM/CSIC-19-124",
    doi = "10.1088/1361-6382/ab5410",
    journal = "Class. Quant. Grav.",
    volume = "37",
    number = "1",
    pages = "015002",
    year = "2020"
}

@article{Buchel:2023fst,
    author = "Buchel, Alex and Cremonini, Sera and Early, Laura",
    title = "{Holographic transport beyond the supergravity approximation}",
    eprint = "2312.05377",
    archivePrefix = "arXiv",
    primaryClass = "hep-th",
    doi = "10.1007/JHEP04(2024)032",
    journal = "JHEP",
    volume = "04",
    pages = "032",
    year = "2024"
}

@article{Maldacena:1997re,
    author = "Maldacena, Juan Martin",
    title = "{The Large $N$ limit of superconformal field theories and supergravity}",
    eprint = "hep-th/9711200",
    archivePrefix = "arXiv",
    reportNumber = "HUTP-97-A097, HUTP-98-A097",
    doi = "10.4310/ATMP.1998.v2.n2.a1",
    journal = "Adv. Theor. Math. Phys.",
    volume = "2",
    pages = "231--252",
    year = "1998"
}

@article{Itzhaki:1998dd,
    author = "Itzhaki, Nissan and Maldacena, Juan Martin and Sonnenschein, Jacob and Yankielowicz, Shimon",
    title = "{Supergravity and the large N limit of theories with sixteen supercharges}",
    eprint = "hep-th/9802042",
    archivePrefix = "arXiv",
    reportNumber = "TAUP-2474-98, HUTP-98-A003",
    doi = "10.1103/PhysRevD.58.046004",
    journal = "Phys. Rev. D",
    volume = "58",
    pages = "046004",
    year = "1998"
}

@article{Gubser:1998bc,
    author = "Gubser, S. S. and Klebanov, Igor R. and Polyakov, Alexander M.",
    title = "{Gauge theory correlators from noncritical string theory}",
    eprint = "hep-th/9802109",
    archivePrefix = "arXiv",
    reportNumber = "PUPT-1767",
    doi = "10.1016/S0370-2693(98)00377-3",
    journal = "Phys. Lett. B",
    volume = "428",
    pages = "105--114",
    year = "1998"
}

@article{Bobev:2021qxx,
    author = "Bobev, Nikolay and Hristov, Kiril and Reys, Valentin",
    title = "{AdS$_{5}$ holography and higher-derivative supergravity}",
    eprint = "2112.06961",
    archivePrefix = "arXiv",
    primaryClass = "hep-th",
    doi = "10.1007/JHEP04(2022)088",
    journal = "JHEP",
    volume = "04",
    pages = "088",
    year = "2022"
}

@article{DeClerck:2023fax,
    author = "De Clerck, Marine and Hartnoll, Sean A. and Santos, Jorge E.",
    title = "{Mixmaster chaos in an AdS black hole interior}",
    eprint = "2312.11622",
    archivePrefix = "arXiv",
    primaryClass = "hep-th",
    doi = "10.1007/JHEP07(2024)202",
    journal = "JHEP",
    volume = "07",
    pages = "202",
    year = "2024"
}

@article{Aharony:2005bm,
    author = "Aharony, Ofer and Minwalla, Shiraz and Wiseman, Toby",
    title = "{Plasma-balls in large N gauge theories and localized black holes}",
    eprint = "hep-th/0507219",
    archivePrefix = "arXiv",
    reportNumber = "WIS-18-05-JUL-DPP, HUTP-05-A0035",
    doi = "10.1088/0264-9381/23/7/001",
    journal = "Class. Quant. Grav.",
    volume = "23",
    pages = "2171--2210",
    year = "2006"
}

@article{Belinsky:1970ew,
    author = "Belinsky, V. A. and Khalatnikov, I. M. and Lifshitz, E. M.",
    title = "{Oscillatory approach to a singular point in the relativistic cosmology}",
    doi = "10.1080/00018737000101171",
    journal = "Adv. Phys.",
    volume = "19",
    pages = "525--573",
    year = "1970"
}

@article{Urbach:2023npi,
    author = "Urbach, Erez Y.",
    title = "{The black hole/string transition in AdS$_{3}$ and confining backgrounds}",
    eprint = "2303.09567",
    archivePrefix = "arXiv",
    primaryClass = "hep-th",
    doi = "10.1007/JHEP09(2023)156",
    journal = "JHEP",
    volume = "09",
    pages = "156",
    year = "2023"
}

@article{Bueno:2025jgc,
    author = "Bueno, Pablo and Lasso Andino, Oscar and Moreno, Javier and van der Velde, Guido",
    title = "{On regular charged black holes in three dimensions}",
    eprint = "2503.02930",
    archivePrefix = "arXiv",
    primaryClass = "gr-qc",
    month = "3",
    year = "2025"
}

@article{Aguayo:2025xfi,
    author = "Aguayo, Monserrat and Gajardo, Leonardo and Grandi, Nicol\'as and Moreno, Javier and Oliva, Julio and Reyes, Mart\'\i{}n",
    title = "{Holographic explorations of regular black holes in pure gravity}",
    eprint = "2505.11736",
    archivePrefix = "arXiv",
    primaryClass = "hep-th",
    month = "6",
    year = "2025"
}

@article{Gubser:2000nd,
    author = "Gubser, Steven S.",
    title = "{Curvature singularities: The Good, the bad, and the naked}",
    eprint = "hep-th/0002160",
    archivePrefix = "arXiv",
    reportNumber = "PUPT-1916",
    doi = "10.4310/ATMP.2000.v4.n3.a6",
    journal = "Adv. Theor. Math. Phys.",
    volume = "4",
    pages = "679--745",
    year = "2000"
}

@article{Meissner:1996sa,
    author = "Meissner, Krzysztof A.",
    title = "{Symmetries of higher order string gravity actions}",
    eprint = "hep-th/9610131",
    archivePrefix = "arXiv",
    reportNumber = "CERN-TH-96-291",
    doi = "10.1016/S0370-2693(96)01556-0",
    journal = "Phys. Lett. B",
    volume = "392",
    pages = "298--304",
    year = "1997"
}

@article{Hohm:2019ccp,
    author = "Hohm, Olaf and Zwiebach, Barton",
    title = "{Non-perturbative de Sitter vacua via $\alpha'$ corrections}",
    eprint = "1905.06583",
    archivePrefix = "arXiv",
    primaryClass = "hep-th",
    doi = "10.1142/S0218271819430028",
    journal = "Int. J. Mod. Phys. D",
    volume = "28",
    number = "14",
    pages = "1943002",
    year = "2019"
}

@article{Veneziano:1991ek,
    author = "Veneziano, G.",
    title = "{Scale factor duality for classical and quantum strings}",
    reportNumber = "CERN-TH-6077-91",
    doi = "10.1016/0370-2693(91)90055-U",
    journal = "Phys. Lett. B",
    volume = "265",
    pages = "287--294",
    year = "1991"
}

@article{Dai:1989ua,
    author = "Dai, Jin and Leigh, R. G. and Polchinski, Joseph",
    title = "{New Connections Between String Theories}",
    reportNumber = "UTTG-12-89",
    doi = "10.1142/S0217732389002331",
    journal = "Mod. Phys. Lett. A",
    volume = "4",
    pages = "2073--2083",
    year = "1989"
}

@article{Dine:1988nrl,
    author = "Dine, Michael and Huet, Patrick Y. and Seiberg, N.",
    title = "{Large and Small Radius in String Theory}",
    reportNumber = "IASSNS-HEP-88/54, CCNY-HEP-88/20",
    doi = "10.1016/0550-3213(89)90418-5",
    journal = "Nucl. Phys. B",
    volume = "322",
    pages = "301--316",
    year = "1989"
}

@article{Li:1998kd,
    author = "Li, Miao",
    title = "{Evidence for large N phase transition in N=4 superYang-Mills theory at finite temperature}",
    eprint = "hep-th/9807196",
    archivePrefix = "arXiv",
    reportNumber = "EFI-98-25",
    doi = "10.1088/1126-6708/1999/03/004",
    journal = "JHEP",
    volume = "03",
    pages = "004",
    year = "1999"
}

@article{Gao:1998ww,
    author = "Gao, Yi-hong and Li, Miao",
    title = "{Large N strong / weak coupling phase transition and the correspondence principle}",
    eprint = "hep-th/9810053",
    archivePrefix = "arXiv",
    reportNumber = "AS-ITP-98-11, EFI-98-49",
    doi = "10.1016/S0550-3213(99)00234-5",
    journal = "Nucl. Phys. B",
    volume = "551",
    pages = "229--241",
    year = "1999"
}

@article{Hartnoll:2005ju,
    author = "Hartnoll, Sean A. and Kumar, S. Prem",
    title = "{AdS black holes and thermal Yang-Mills correlators}",
    eprint = "hep-th/0508092",
    archivePrefix = "arXiv",
    reportNumber = "DAMTP-2005-73",
    doi = "10.1088/1126-6708/2005/12/036",
    journal = "JHEP",
    volume = "12",
    pages = "036",
    year = "2005"
}

@article{Bergshoeff:1995as,
    author = "Bergshoeff, Eric and Hull, Christopher M. and Ortin, Tomas",
    title = "{Duality in the type II superstring effective action}",
    eprint = "hep-th/9504081",
    archivePrefix = "arXiv",
    reportNumber = "UG-3-95, QMW-PH-95-4",
    doi = "10.1016/0550-3213(95)00367-2",
    journal = "Nucl. Phys. B",
    volume = "451",
    pages = "547--578",
    year = "1995"
}

@article{Giveon:1991jj,
    author = "Giveon, Amit and Rocek, Martin",
    title = "{Generalized duality in curved string backgrounds}",
    eprint = "hep-th/9112070",
    archivePrefix = "arXiv",
    reportNumber = "IASSNS-HEP-91-84, ITP-SB-91-67",
    doi = "10.1016/0550-3213(92)90518-G",
    journal = "Nucl. Phys. B",
    volume = "380",
    pages = "128--146",
    year = "1992"
}

@article{Giveon:1994fu,
    author = "Giveon, Amit and Porrati, Massimo and Rabinovici, Eliezer",
    title = "{Target space duality in string theory}",
    eprint = "hep-th/9401139",
    archivePrefix = "arXiv",
    reportNumber = "RI-1-94, NYU-TH-94-01-01",
    doi = "10.1016/0370-1573(94)90070-1",
    journal = "Phys. Rept.",
    volume = "244",
    pages = "77--202",
    year = "1994"
}

@article{Hsia:2024kpi,
    author = "Hsia, Steven Weilong and Kamal, Ahmed Rakin and Wulff, Linus",
    title = "{No manifest T duality at order {\ensuremath{\alpha}}'3}",
    eprint = "2411.15302",
    archivePrefix = "arXiv",
    primaryClass = "hep-th",
    doi = "10.1103/PhysRevD.111.L061904",
    journal = "Phys. Rev. D",
    volume = "111",
    number = "6",
    pages = "L061904",
    year = "2025"
}

@article{Wu:2024eci,
    author = "Wu, Houwen and Yan, Zihan and Ying, Shuxuan",
    title = "{Revisiting Schwarzschild black hole singularity through string theory}",
    eprint = "2402.05870",
    archivePrefix = "arXiv",
    primaryClass = "hep-th",
    month = "2",
    year = "2024"
}

@article{Codina:2023nwz,
    author = "Codina, Tomas and Hohm, Olaf and Zwiebach, Barton",
    title = "{Black hole singularity resolution in D=2 via duality-invariant \ensuremath{\alpha}' corrections}",
    eprint = "2308.09743",
    archivePrefix = "arXiv",
    primaryClass = "hep-th",
    reportNumber = "HU-EP-23/46-RTG, MIT-CTP/5595",
    doi = "10.1103/PhysRevD.108.126006",
    journal = "Phys. Rev. D",
    volume = "108",
    number = "12",
    pages = "126006",
    year = "2023"
}

@article{Caceres:2024edr,
    author = "C{\'a}ceres, Elena and Murcia, {\'A}ngel J. and Patra, Ayan K. and Pedraza, Juan F.",
    title = "{Kasner eons with matter: holographic excursions to the black hole singularity}",
    eprint = "2408.14535",
    archivePrefix = "arXiv",
    primaryClass = "hep-th",
    reportNumber = "WI-27-2024, IFT-UAM/CSIC-24-123",
    doi = "10.1007/JHEP12(2024)077",
    journal = "JHEP",
    volume = "12",
    pages = "077",
    year = "2024"
}

@article{Gross:1986iv,
    author = "Gross, David J. and Witten, Edward",
    title = "{Superstring Modifications of Einstein's Equations}",
    reportNumber = "Print-86-0250 (PRINCETON)",
    doi = "10.1016/0550-3213(86)90429-3",
    journal = "Nucl. Phys. B",
    volume = "277",
    pages = "1",
    year = "1986"
}

@article{Meissner:1991zj,
    author = "Meissner, K. A. and Veneziano, G.",
    title = "{Symmetries of cosmological superstring vacua}",
    reportNumber = "CERN-TH-6138-91",
    doi = "10.1016/0370-2693(91)90520-Z",
    journal = "Phys. Lett. B",
    volume = "267",
    pages = "33--36",
    year = "1991"
}

@article{Nunez:2020hxx,
    author = "N\'u\~nez, Carmen A. and Rost, Facundo Emanuel",
    title = "{New non-perturbative de Sitter vacua in $\alpha'$-complete cosmology}",
    eprint = "2011.10091",
    archivePrefix = "arXiv",
    primaryClass = "hep-th",
    doi = "10.1007/JHEP03(2021)007",
    journal = "JHEP",
    volume = "03",
    pages = "007",
    year = "2021"
}

@article{Dijkgraaf:1991ba,
    author = "Dijkgraaf, Robbert and Verlinde, Herman L. and Verlinde, Erik P.",
    title = "{String propagation in a black hole geometry}",
    reportNumber = "PUPT-1252, IASSNS-HEP-91-22",
    doi = "10.1016/0550-3213(92)90237-6",
    journal = "Nucl. Phys. B",
    volume = "371",
    pages = "269--314",
    year = "1992"
}

@article{Witten:1991yr,
    author = "Witten, Edward",
    title = "{On string theory and black holes}",
    reportNumber = "IASSNS-HEP-91-12",
    doi = "10.1103/PhysRevD.44.314",
    journal = "Phys. Rev. D",
    volume = "44",
    pages = "314--324",
    year = "1991"
}

@inproceedings{DHoker:2002nbb,
    author = "D'Hoker, Eric and Freedman, Daniel Z.",
    title = "{Supersymmetric gauge theories and the AdS / CFT correspondence}",
    booktitle = "{Theoretical Advanced Study Institute in Elementary Particle Physics (TASI 2001): Strings, Branes and EXTRA Dimensions}",
    eprint = "hep-th/0201253",
    archivePrefix = "arXiv",
    reportNumber = "UCLA-02-TEP-3, MIT-CTP-3242",
    pages = "3--158",
    month = "1",
    year = "2002"
}

@article{Witten:1998qj,
    author = "Witten, Edward",
    title = "{Anti de Sitter space and holography}",
    eprint = "hep-th/9802150",
    archivePrefix = "arXiv",
    reportNumber = "IASSNS-HEP-98-15",
    doi = "10.4310/ATMP.1998.v2.n2.a2",
    journal = "Adv. Theor. Math. Phys.",
    volume = "2",
    pages = "253--291",
    year = "1998"
}

@article{Codina:2020kvj,
    author = "Codina, Tomas and Hohm, Olaf and Marques, Diego",
    title = "{String Dualities at Order $\alpha'^{\,3}$}",
    eprint = "2012.15677",
    archivePrefix = "arXiv",
    primaryClass = "hep-th",
    reportNumber = "HU-EP-20/46-RTG",
    doi = "10.1103/PhysRevLett.126.171602",
    journal = "Phys. Rev. Lett.",
    volume = "126",
    number = "17",
    pages = "171602",
    year = "2021"
}

@article{Festuccia:2005pi,
    author = "Festuccia, Guido and Liu, Hong",
    title = "{Excursions beyond the horizon: Black hole singularities in Yang-Mills theories. I.}",
    eprint = "hep-th/0506202",
    archivePrefix = "arXiv",
    reportNumber = "MIT-CTP-3641",
    doi = "10.1088/1126-6708/2006/04/044",
    journal = "JHEP",
    volume = "04",
    pages = "044",
    year = "2006"
}

@article{Codina:2021cxh,
    author = "Codina, Tomas and Hohm, Olaf and Marques, Diego",
    title = "{General string cosmologies at order \ensuremath{\alpha}'3}",
    eprint = "2107.00053",
    archivePrefix = "arXiv",
    primaryClass = "hep-th",
    reportNumber = "HU-EP-21/16-RTG",
    doi = "10.1103/PhysRevD.104.106007",
    journal = "Phys. Rev. D",
    volume = "104",
    number = "10",
    pages = "106007",
    year = "2021"
}

@article{Hohm:2019jgu,
    author = "Hohm, Olaf and Zwiebach, Barton",
    title = "{Duality invariant cosmology to all orders in $\alpha$'}",
    eprint = "1905.06963",
    archivePrefix = "arXiv",
    primaryClass = "hep-th",
    reportNumber = "MIT-CTP-5119, HU-EP-19/07",
    doi = "10.1103/PhysRevD.100.126011",
    journal = "Phys. Rev. D",
    volume = "100",
    number = "12",
    pages = "126011",
    year = "2019"
}

@article{Sen:1991zi,
    author = "Sen, Ashoke",
    title = "{O(d) x O(d) symmetry of the space of cosmological solutions in string theory, scale factor duality and two-dimensional black holes}",
    reportNumber = "TIFR-TH-91-35",
    doi = "10.1016/0370-2693(91)90090-D",
    journal = "Phys. Lett. B",
    volume = "271",
    pages = "295--300",
    year = "1991"
}

@article{Gursoy:2007cb,
    author = "Gursoy, U. and Kiritsis, E.",
    title = "{Exploring improved holographic theories for QCD: Part I}",
    eprint = "0707.1324",
    archivePrefix = "arXiv",
    primaryClass = "hep-th",
    reportNumber = "CPHT-RR027-0507",
    doi = "10.1088/1126-6708/2008/02/032",
    journal = "JHEP",
    volume = "02",
    pages = "032",
    year = "2008"
}

@article{Gursoy:2007er,
    author = "Gursoy, U. and Kiritsis, E. and Nitti, F.",
    title = "{Exploring improved holographic theories for QCD: Part II}",
    eprint = "0707.1349",
    archivePrefix = "arXiv",
    primaryClass = "hep-th",
    reportNumber = "CPHT-RR028-0507",
    doi = "10.1088/1126-6708/2008/02/019",
    journal = "JHEP",
    volume = "02",
    pages = "019",
    year = "2008"
}

@article{Frenkel:2020ysx,
    author = "Frenkel, Alexander and Hartnoll, Sean A. and Kruthoff, Jorrit and Shi, Zhengyan D.",
    title = "{Holographic flows from CFT to the Kasner universe}",
    eprint = "2004.01192",
    archivePrefix = "arXiv",
    primaryClass = "hep-th",
    doi = "10.1007/JHEP08(2020)003",
    journal = "JHEP",
    volume = "08",
    pages = "003",
    year = "2020"
}

@article{Caceres:2022smh,
    author = "Caceres, Elena and Kundu, Arnab and Patra, Ayan K. and Shashi, Sanjit",
    title = "{Trans-IR flows to black hole singularities}",
    eprint = "2201.06579",
    archivePrefix = "arXiv",
    primaryClass = "hep-th",
    reportNumber = "UTTG-30-2022",
    doi = "10.1103/PhysRevD.106.046005",
    journal = "Phys. Rev. D",
    volume = "106",
    number = "4",
    pages = "046005",
    year = "2022"
}

@article{Caceres:2023zhl,
    author = "Caceres, Elena and Shashi, Sanjit and Sun, Hao-Yu",
    title = "{Imprints of phase transitions on Kasner singularities}",
    eprint = "2305.11177",
    archivePrefix = "arXiv",
    primaryClass = "hep-th",
    reportNumber = "UTWI-17-2023",
    doi = "10.1103/PhysRevD.109.126018",
    journal = "Phys. Rev. D",
    volume = "109",
    number = "12",
    pages = "126018",
    year = "2024"
}

@article{Caceres:2023zft,
    author = "C{\'a}ceres, Elena and Patra, Ayan K. and Pedraza, Juan F.",
    title = "{Shock waves, black hole interiors and holographic RG flows}",
    eprint = "2311.12940",
    archivePrefix = "arXiv",
    primaryClass = "hep-th",
    reportNumber = "UTWI-42-2023, IFT-UAM/CSIC-23-148",
    doi = "10.1007/JHEP07(2024)052",
    journal = "JHEP",
    volume = "07",
    pages = "052",
    year = "2024"
}

@article{Bueno:2024fzg,
    author = "Bueno, Pablo and Cano, Pablo A. and Hennigar, Robie A.",
    title = "{Kasner epochs, eras and eons}",
    eprint = "2402.14912",
    archivePrefix = "arXiv",
    primaryClass = "gr-qc",
    doi = "10.1103/PhysRevD.110.L041503",
    journal = "Phys. Rev. D",
    volume = "110",
    number = "4",
    pages = "L041503",
    year = "2024"
}

@article{Bueno:2019ltp,
    author = "Bueno, Pablo and Cano, Pablo A. and Moreno, Javier and Murcia, {\'A}ngel",
    title = "{All higher-curvature gravities as Generalized quasi-topological gravities}",
    eprint = "1906.00987",
    archivePrefix = "arXiv",
    primaryClass = "hep-th",
    doi = "10.1007/JHEP11(2019)062",
    journal = "JHEP",
    volume = "11",
    pages = "062",
    year = "2019"
}

@article{Beisert:2010jr,
    author = "Beisert, Niklas and others",
    title = "{Review of AdS/CFT Integrability: An Overview}",
    eprint = "1012.3982",
    archivePrefix = "arXiv",
    primaryClass = "hep-th",
    reportNumber = "AEI-2010-175, CERN-PH-TH-2010-306, HU-EP-10-87, HU-MATH-2010-22, KCL-MTH-10-10, UMTG-270, UUITP-41-10",
    doi = "10.1007/s11005-011-0529-2",
    journal = "Lett. Math. Phys.",
    volume = "99",
    pages = "3--32",
    year = "2012"
}

@article{Gaberdiel:2021qbb,
    author = "Gaberdiel, Matthias R. and Gopakumar, Rajesh",
    title = "{String Dual to Free N=4 Supersymmetric Yang-Mills Theory}",
    eprint = "2104.08263",
    archivePrefix = "arXiv",
    primaryClass = "hep-th",
    doi = "10.1103/PhysRevLett.127.131601",
    journal = "Phys. Rev. Lett.",
    volume = "127",
    number = "13",
    pages = "131601",
    year = "2021"
}

@article{Freedman:1999gp,
    author = "Freedman, D. Z. and Gubser, S. S. and Pilch, K. and Warner, N. P.",
    title = "{Renormalization group flows from holography supersymmetry and a c theorem}",
    eprint = "hep-th/9904017",
    archivePrefix = "arXiv",
    reportNumber = "CERN-TH-99-86, HUTP-99-A015, USC-99-1, MIT-CTP-2846",
    doi = "10.4310/ATMP.1999.v3.n2.a7",
    journal = "Adv. Theor. Math. Phys.",
    volume = "3",
    pages = "363--417",
    year = "1999"
}

@article{Kiritsis:2009hu,
    author = "Kiritsis, Elias",
    editor = "Lust, Dieter and Dobrev, Vladimir",
    title = "{Dissecting the string theory dual of QCD}",
    eprint = "0901.1772",
    archivePrefix = "arXiv",
    primaryClass = "hep-th",
    doi = "10.1002/prop.200900011",
    journal = "Fortsch. Phys.",
    volume = "57",
    pages = "396--417",
    year = "2009"
}

@article{Ying:2022xaj,
    author = "Ying, Shuxuan",
    title = "{Two-dimensional regular string black hole via complete $\alpha ^{\prime }$ corrections}",
    eprint = "2212.03808",
    archivePrefix = "arXiv",
    primaryClass = "hep-th",
    doi = "10.1140/epjc/s10052-023-11756-9",
    journal = "Eur. Phys. J. C",
    volume = "83",
    number = "7",
    pages = "577",
    year = "2023"
}

@article{Ying:2021xse,
    author = "Ying, Shuxuan",
    title = "{Resolving naked singularities in $\alpha ^{\prime }$-corrected string theory}",
    eprint = "2112.03087",
    archivePrefix = "arXiv",
    primaryClass = "hep-th",
    doi = "10.1140/epjc/s10052-022-10427-5",
    journal = "Eur. Phys. J. C",
    volume = "82",
    number = "6",
    pages = "523",
    year = "2022"
}

@article{Moitra:2025fhx,
    author = "Moitra, Upamanyu",
    title = "{Heterotic Black Holes in Duality-Invariant Formalism}",
    eprint = "2511.21687",
    archivePrefix = "arXiv",
    primaryClass = "hep-th",
    month = "11",
    year = "2025"
}

@article{Buchel:2026xtn,
    author = "Buchel, Alex and Cremonini, Sera and Moezzi, Mohammad and Tringas, George",
    title = "{Holographic Charged Transport with Higher Derivatives}",
    eprint = "2602.06144",
    archivePrefix = "arXiv",
    primaryClass = "hep-th",
    month = "2",
    year = "2026"
}

@article{Apostolidis:2025gnn,
    author = {Apostolidis, Thomas and G{\"u}rsoy, Umut and Pr{\'e}au, Edwan},
    title = "{Higher derivative holography and temperature dependence of QGP viscosities}",
    eprint = "2502.19195",
    archivePrefix = "arXiv",
    primaryClass = "hep-th",
    doi = "10.1007/JHEP11(2025)131",
    journal = "JHEP",
    volume = "11",
    pages = "131",
    year = "2025"
}

@article{Polyakov:1998ju,
    author = "Polyakov, Alexander M.",
    title = "{The Wall of the cave}",
    eprint = "hep-th/9809057",
    archivePrefix = "arXiv",
    reportNumber = "PUPT-1812",
    doi = "10.1142/S0217751X99000324",
    journal = "Int. J. Mod. Phys. A",
    volume = "14",
    pages = "645--658",
    year = "1999"
}

@article{Skenderis:1999mm,
    author = "Skenderis, Kostas and Townsend, Paul K.",
    title = "{Gravitational stability and renormalization group flow}",
    eprint = "hep-th/9909070",
    archivePrefix = "arXiv",
    reportNumber = "DAMTP-1999-111, SPIN-1999-20",
    doi = "10.1016/S0370-2693(99)01212-5",
    journal = "Phys. Lett. B",
    volume = "468",
    pages = "46--51",
    year = "1999"
}

@article{Kiritsis:2014kua,
    author = "Kiritsis, Elias and Li, Wenliang and Nitti, Francesco",
    title = "{Holographic RG flow and the Quantum Effective Action}",
    eprint = "1401.0888",
    archivePrefix = "arXiv",
    primaryClass = "hep-th",
    reportNumber = "CERN-PH-TH-2013-321",
    doi = "10.1002/prop.201400007",
    journal = "Fortsch. Phys.",
    volume = "62",
    pages = "389--454",
    year = "2014"
}

@article{Hartnoll:2020fhc,
    author = "Hartnoll, Sean A. and Horowitz, Gary T. and Kruthoff, Jorrit and Santos, Jorge E.",
    title = "{Diving into a holographic superconductor}",
    eprint = "2008.12786",
    archivePrefix = "arXiv",
    primaryClass = "hep-th",
    doi = "10.21468/SciPostPhys.10.1.009",
    journal = "SciPost Phys.",
    volume = "10",
    number = "1",
    pages = "009",
    year = "2021"
}

@article{Cai:2020wrp,
    author = "Cai, Rong-Gen and Li, Li and Yang, Run-Qiu",
    title = "{No Inner-Horizon Theorem for Black Holes with Charged Scalar Hairs}",
    eprint = "2009.05520",
    archivePrefix = "arXiv",
    primaryClass = "gr-qc",
    doi = "10.1007/JHEP03(2021)263",
    journal = "JHEP",
    volume = "03",
    pages = "263",
    year = "2021"
}

@article{VandeMoortel:2021gsp,
    author = "Van de Moortel, Maxime",
    title = "{Violent Nonlinear Collapse in the Interior of Charged Hairy Black Holes}",
    eprint = "2109.10932",
    archivePrefix = "arXiv",
    primaryClass = "gr-qc",
    doi = "10.1007/s00205-024-02038-z",
    journal = "Arch. Ration. Mech. Anal.",
    volume = "248",
    number = "5",
    pages = "89",
    year = "2024"
}

@article{Li:2023tfa,
    author = "Li, Warren and Van de Moortel, Maxime",
    title = "{Kasner Bounces and Fluctuating Collapse Inside Hairy Black Holes with Charged Matter}",
    eprint = "2302.00046",
    archivePrefix = "arXiv",
    primaryClass = "gr-qc",
    doi = "10.1007/s40818-024-00192-x",
    journal = "Ann. PDE",
    volume = "11",
    number = "1",
    pages = "3",
    year = "2025"
}

\end{document}